%% file: talk.tex
\documentclass[12pt]{article}
\usepackage{amssymb,axodraw}
\begin{document}

\renewcommand{\thefootnote}{\fnsymbol{footnote}}

\mbox{ } \\[-1cm]
\mbox{ }\hfill TUM--HEP--488/02\\%[-1mm] 
\mbox{ }\hfill hep--ph/0210280\\%[-1mm] 
\mbox{ }\hfill \today\\%[-1mm]

\begin{center}
  {\Large\bf Ultra-high energy cosmic rays from super-heavy X
    particle decay} \\[8mm]
           
Cyrille Barbot \\[4mm]

{\it Physik Dept., TU M\"unchen, James Franck Str., D--85748
 Garching, Germany} \\[1mm]
\end{center}

\bigskip
\bigskip 
\bigskip 

\begin{abstract}

\vskip 0.5cm

\noindent
%PACS number(s): 11.30.Pb, 11.30.Er

In this talk, I present the last and more precise results obtained in
the computation of the final spectra of stable particles issued from
the decay of super-heavy X particles ($M_X \sim 10^{21}$ to $10^{25}$
eV). Such very energetic decay products, carrying a fraction of the
mass of the X particle, are believed to be a plausible explanation for
the observed ultra-high energy cosmic rays (UHECR). Combining these
results with X-particle models and with a code describing the
propagation effects for UHECRs through the interstellar medium, it
becomes possible to make some predictions on the fluxes expected on
Earth, hopefully detectable in the next generation of experiments.

\end{abstract}

\setcounter{footnote}{0}

In the second part of the 20th century, the spectrum of cosmic rays
(CRs) has been mesured over more than 12 decades of energy. Even if
our understanding of it has grown a lot in the last few decades, many
enigma are remaining. One of them concerns the extremity of this
spectrum, at the highest energies, where theorists were expecting a
strong cut-off to occur at energies of the order of $5.10^{19}$ eV :
indeed, at these energies, CRs should be of extragalactic origin, and
probably coming from distances further than the local cluster of
galaxies, because we know no astrophysical object able to accelerate
particles enough to give them this energy in our vicinity. But the
point is that particles carrying energies above $10^{20}$ eV traveling
over cosmological distances should loose their energy through propagation
effects ; for example, a proton will interact with the cosmological
microwave background (CMB) and photoproduce pions, with an interaction
length of a few tens of Mpc, loosing around 20 \% of its energy at
each interaction. Similar processes occur with nuclei, photons or
electrons. Thus particles with initial energy $\sim 10^{20}$ eV should reach
the Earth with a maximal energy $\sim 5.10^{19}$ eV, the so-called GZK
cut-off \cite{GZK:1,GZK:2}.\footnote{A notable exception are, of
  course, the neutrinos, which can travel over cosmological distances
  without loosing their energy. But the events observed on Earth
  cannot be attributed to primary neutrinos.} The fact is that events
have been registered above this cut-off in very different experiments
over the last few decades \cite{HaverahPark,FlysEye,HIRES,AGASA}. Such
an observation is almost impossible to reconcile with any model of
acceleration of charged particles in any astrophysical object.
Moreover, there is another strong indication against these models :
UHECRs are expected to travel rather straight away
in the universe, without being deviated by the (inter)galactic
magnetic fields. Thus they should point to there sources within a few
degrees. Yet, excepted the existence of a few doublets and triplets in
the experimental data, the observations are compatible with an almost
perfect isotropy \cite{Anisotropy}\footnote{Nevertheless, it should be
  noted that there are still attempts to explain the UHECRs with these
  classical ``bottom-up'' theories, see for example
  \cite{WaxmanGRB,Biermann}.}.

These remarks lead to the development of another class of models for
explaining the existence of UHECRs, namely the ``top-down'' theories,
which are considering that the observed events could be generated
through the decay of some mysterious super-heavy ``X'' particles. The
existence of such X particles is predicted in number of GUT theories
or in relation with topological defects collapsing or annihilating,
and they can be created rather naturally at the end of the inflation
\cite{Allahverdi:1,Allahverdi:2}.  Among other more ``model
dependant'' properties, top-down models require that these particles
should have a mass bigger than the highest energies observed in UHECR
events, $M_X > 10^{21}$ eV, and a lifetime of the order of (or greater
than) the age of the universe.  They could be trapped homogeneously in
the galaxies\footnote{or possibly in some topological defects
  protecting them against decay, with distributions very different
  from matter distributions in the galaxies.}, explaining the isotropy
of the data, and would constitute semi-local sources for UHECRs,
avoiding the GZK problem.  Moreover, if they are abundant enough and
trapped in the galaxies, they could be a very good candidate for the
dark matter problem \cite{SHDM:1}.

Excellent reviews on the UHECRs can be found in the litterature
\cite{reviewSigl,reviewSarkar,reviewStecker}. In this talk I will
focus on the top-down theories, and give general results for the decay
of ultra-heavy particles (first presented in \cite{BarbotDrees:1}),
independantly of any particular model.\newline

\setcounter{footnote}{0}

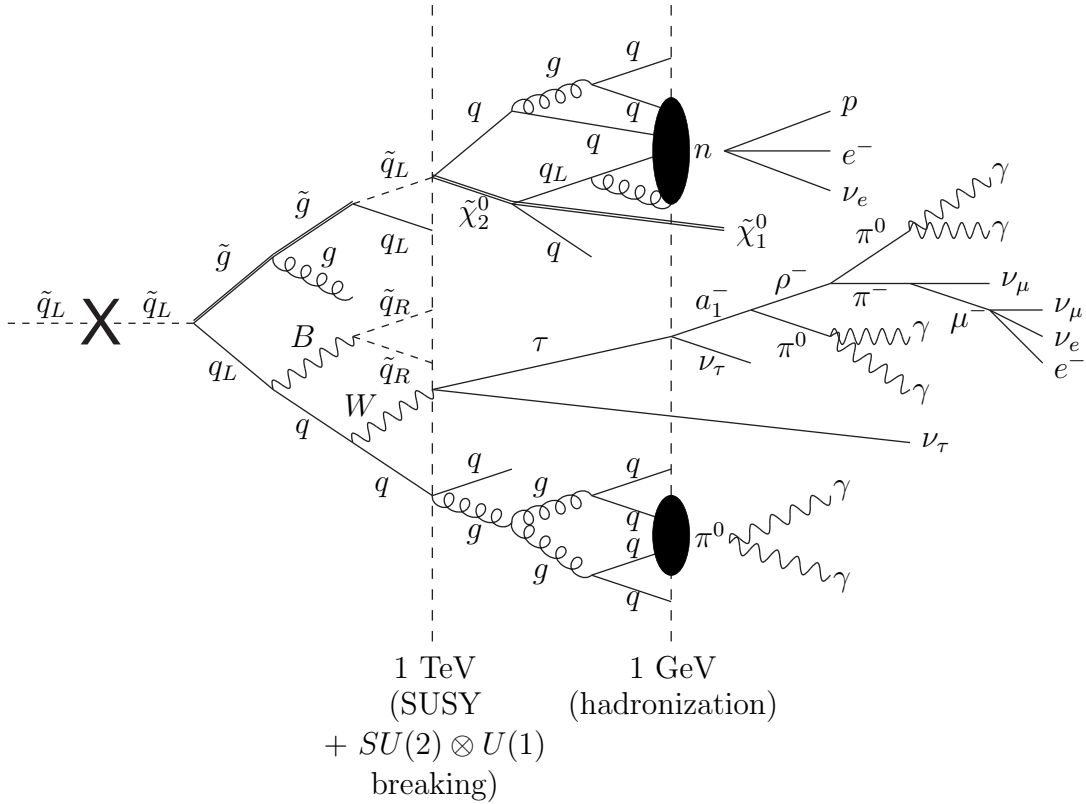
\begin{figure}[h!]
\begin{center} \begin{picture}(-400,300)(0,-180)
\SetPFont{Helvetica}{24}
\PText(-405,4)(0)[]{X}
\DashLine(-410,0)(-440,0){3} \Text(-425,7)[]{$\tilde{q}_L$} % squark 
\DashLine(-400,0)(-370,0){3} \Text(-385,7)[]{$\tilde{q}_L$} % squark 
\Line(-370,1)(-340,26)
\Line(-370,0)(-340,25) \Text(-360,25)[]{$\tilde{g}$} % gluino
\Line(-340,26)(-310,46)
\Line(-340,25)(-310,45) \Text(-330,45)[]{$\tilde{g}$}% gluino
\Gluon(-340,25)(-310,10){-3}{4}\Text(-320,25)[]{$g$} % gluon
\DashLine(-310,45)(-280,55){3}  \Text(-295,60)[]{$\tilde{q}_L$} %squark
\Line(-310,45)(-280,35) \Text(-295,30)[]{$q_L$} % quark
\DashLine(-280,-120)(-280,120){4} \Text(-280,-130)[c]{1 TeV}% 1 TeV
\Text(-280,-145)[c]{(SUSY}
\Text(-280,-160)[c]{+ $SU(2)\otimes U(1)$}
\Text(-280,-175)[c] {breaking)}

\Line(-280,55)(-250,80) \Text(-265,80)[]{$q$} % quark
\Line(-250,80)(-190,70) \Text(-220,68)[]{$q$} % quark
\Gluon(-250,80)(-220,90){3}{4} \Text(-235,97)[]{$g$} % gluon
\Line(-220,90)(-190,100) \Text(-205,103)[]{$q$} % quark
\Line(-220,90)(-190,80) \Text(-205,80)[]{$q$} % quark

\Line(-280,56)(-250,46)
\Line(-280,55)(-250,45) \Text(-265,42)[]{$\tilde{\chi}_2^0$}%neutralino 2 
\Line(-250,45)(-220,55) \Text(-235,57)[]{$q_L$}% quark
\Line(-250,46)(-170,36)
\Line(-250,45)(-170,35) \Text(-166,35)[l]{$\tilde{\chi}_1^0$}% neutralino 1 
\Line(-250,45)(-220,25) \Text(-235,27)[]{$q$} % quark
\Line(-220,55)(-190,65) % \Text(-205,65)[]{$q_L$}% quark
\Gluon(-220,55)(-190,45){-3}{4} %gluon
\DashLine(-190,-120)(-190,120){4} \Text(-190,-130)[c]{1 GeV}% 1 GeV
\Text(-190,-145)[c]{(hadronization)}

\Line(-370,0)(-340,-25) \Text(-360,-18)[]{$q_L$} % quark
\Photon(-340,-25)(-310,-5){3}{5} \Text(-330,-5)[]{$B$} % B
\DashLine(-310,-5)(-280,5){3} \Text(-295,8)[]{$\tilde{q}_R$}% squark
\DashLine(-310,-5)(-280,-15){3} \Text(-295,-18)[]{$\tilde{q}_R$}% squark
\Line(-340,-25)(-310,-45) \Text(-330,-40)[]{$q$} % quark
\Photon(-310,-45)(-280,-25){3}{5} \Text(-308,-31)[]{$W$} % W
\Line(-280,-25)(-190,-5) \Text(-240,-8)[]{$\tau$} %tau
\Line(-190,-5)(-160,5) \Text(-175,9)[]{$a_1^{-}$}% a1-
\Line(-160,5)(-130,15) \Text(-145,17)[]{$\rho^{-}$}% rho-
\Line(-130,15)(-100,15) \Text(-115,11)[]{$\pi^{-}$}% pi-
\Line(-100,15)(-70,15) \Text(-66,15)[l]{$\nu_\mu$} %nu_mu
\Line(-100,15)(-70,5) \Text(-78,2)[]{$\mu^{-}$} %mu
\Line(-70,5)(-50,5) \Text(-46,5)[l]{$\nu_\mu$} %nu_mu
\Line(-70,5)(-50,-5) \Text(-46,-7)[l]{$\nu_e$} %nu_e
\Line(-70,5)(-50,-15) \Text(-46,-16)[l]{$e^{-}$} %e

\Line(-130,15)(-100,35) \Text(-115,35)[]{$\pi^0$} % pi0
\Photon(-100,35)(-70,35){3}{5} \Text(-66,35)[]{$\gamma$} % gamma
\Photon(-100,35)(-70,55){3}{5} \Text(-66,57)[]{$\gamma$} % gamma

\Line(-160,5)(-130,-5) \Text(-145,-8)[]{$\pi^0$}% pi0
\Photon(-130,-5)(-100,-25){3}{5} \Text(-96,-27)[]{$\gamma$} % gamma
\Photon(-130,-5)(-100,-5){3}{5} \Text(-96,-3)[]{$\gamma$} % gamma
\Line(-190,-5)(-160,-15) \Text(-175,-16)[]{$\nu_\tau$} % nu_tau
\Line(-280,-25)(-100,-45) \Text(-96,-45)[l]{$\nu_\tau$} %nu_tau

\Line(-310,-45)(-280,-65) \Text(-300,-62)[]{$q$} % quark
\Line(-280,-65)(-250,-55) \Text(-265,-53)[]{$q$} % quark
\Gluon(-280,-65)(-250,-75){-3}{4} \Text(-265,-80)[]{$g$}% gluon
\Gluon(-250,-75)(-220,-95){-3}{4} \Text(-240,-95)[]{$g$} % gluon
\Line(-220,-95)(-190,-105) \Text(-205,-105)[]{$q$} % quark
\Line(-220,-95)(-190,-85) \Text(-205,-85)[]{$q$}% quark

\Gluon(-250,-75)(-220,-65){3}{4} \Text(-240,-62)[]{$g$} % gluon
\Line(-220,-65)(-190,-55)  \Text(-205,-55)[]{$q$}% quark
\Line(-220,-65)(-190,-75)  \Text(-205,-75)[]{$q$}% quark

\GOval(-190,65)(20,7)(0){0} % neutron
\Text(-175,65)[r]{$n$}
\Line(-170,65)(-130,80) \Text(-126,82)[l]{$p$}% proton
\Line(-170,65)(-130,65) \Text(-126,65)[l]{$e^{-}$}% electron
\Line(-170,65)(-130,50) \Text(-126,48)[l]{$\nu_e$} % nu_e

\GOval(-190,-80)(15,7)(0){0} % pi_0
\Text(-170,-80)[r]{$\pi^0$}
\Photon(-168,-83)(-130,-65){3}{5} \Text(-126,-63)[]{$\gamma$} % gamma
\Photon(-168,-83)(-130,-95){3}{5} \Text(-126,-98)[]{$\gamma$} % gamma
\end{picture} 
\caption{Schematic MSSM cascade for an initial squark with
  a virtuality $Q = M_X$. The full circles indicate decays of massive
  particles, in distinction to fragmentation vertices. See the text
  for further details.}
\end{center}
\label{Cascade}
\end{figure}

We first briefly describe the physical steps involved in the decay
cascade of an ultra-heavy X particle in the framework of the MSSM, as
they are illustrated on fig~\ref{cascade}.  Our basic assumption is
that the X particle decays in N very virtual particles of the
MSSM\footnote{The existence of an energy scale as high as $M_X$
  strongly suggests the existence of superparticles with masses not
  much above 1 TeV, in order to guarantee the perturbative stability
  of the hierarchy between $M_X$ and the weak scale. We therefore
  usually allow superparticles as well as ordinary particles to be
  produced in $X$ decays, as described by the minimal supersymmetric
  extension of the Standard Model (MSSM).}, each of them initiating a
decay cascade, following the known physics at lower energy. At high
virtuality, in the region of asymptotic freedom, each of the primaries
will initiate a perturbative shower, splitting into two allowed
particles of smaller virtuality, according to the Feynman laws
\footnote{In contrast to previous works
  \cite{Berezinsky:2000,Coriano:2001,Sarkar:2001,ToldraLSP},
  we considered in our treatment all gauge interactions as well as
  third generation Yukawa interactions, rather than only SUSY--QCD;
  note that at energies above $10^{20}$ eV all gauge interactions are
  of comparable strength. The inclusion of electroweak gauge
  interactions in the shower gives rise to a significant flux of very
  energetic photons and leptons, which had not been identified in
  earlier studies.}. These products will split at their turn too, and
the process will continue until the virtuality has decreased enough,
at a scale where both SUSY and $SU(2) \otimes U(1)$ will break (for
simplicity we are considering a unique SUSY mass scale $M_{SUSY} \sim
1$ TeV for all sparticles). $M_{SUSY}$ is symbolized by the first
vertical dash-line in fig 1. All the on-shell massive
sparticles produced at this stage will then decay into Standard Model
(SM) particles and the only (eventually) stable sparticle, the
so-called Lightest Supersymmetric Particle (LSP). The heavy SM
particles, like the top quarks and the massive bosons, will decay too,
but the lighter quarks and gluons will continue a perturbative
partonic shower until they have reached either their on-shell mass
scale or the typical scale of hadronization (say 1 GeV), the second
vertical dash-line of fig 1.  At this stage, the color
effects become too strong and the partons cannot propagate freely
anymore, being forced to combine into colorless hadronic states.
Finally, the unstable hadrons will also decay, and only the stable
particles will remain and propagate in the intergalactic space, namely
the protons, photons, electrons, the three species of neutrinos and
the LSP (and their antiparticles).

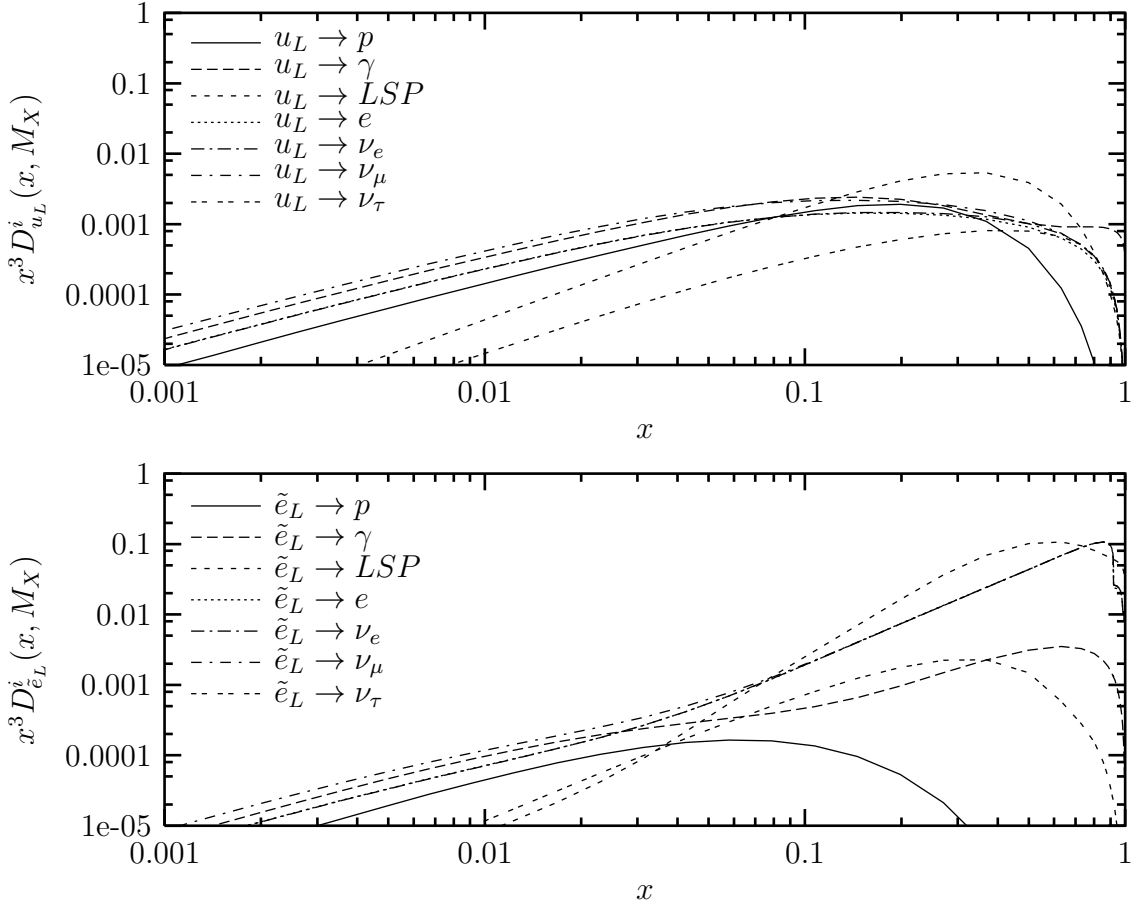
\begin{figure}[h!]
\setlength{\unitlength}{0.3 cm}
\begin{minipage}[t]{4.5cm}
\input{Low_G_uL.tex}
\end{minipage}

\noindent
\begin{minipage}[b]{4.5cm}
\input{Low_G_eL_.tex}
\end{minipage} \hfill
\caption{$x^3 \times$ Fragmentation functions of a first generation SU(2) doublet quark (top) and a slepton (bottom) into stable particles. }
\label{FFs}
\end{figure}

The perturbative part of the shower is treated through the numerical
resolution of DGLAP evolution equations \cite{AP} extended to the
complete spectrum of the MSSM. These equations describe the evolution
of the so-called ``fragmentation functions'' (FFs), which are
describing the fragmentation of any fondamental particle into any
other ; the DGLAP evolution equations describe more specifically the
impact of all 3-legs MSSM Feynman diagrams on these FFs, and
their evolution with energy through the running of the associated
coupling constants. We worked out all the FFs of the MSSM by solving
these equations for all the unbroken fields between $M_{SUSY}$ and
$M_X$. At the breaking scale $M_{SUSY}$ we applied the canonical
unitary transformation to the FFs of the unbroken fields in order to
obtain those of the broken ones, and computed the decay cascade of the
supersymmetric part of the spectrum. We used here the results of the
public code Isasusy \cite{Isasusy} to describe the allowed decays and
their branching ratios, for a given set of SUSY parameters. If
R-parity is conserved, we obtain the final spectrum of the stable LSP
at this step, and the rest of the available energy is distributed
between the SM particles. After a longer perturbative cascade down to
$Q \sim \max(m_{quark},1 GeV)$, as stated before, the quarks and
gluons will hadronize. We used the results of \cite{Poetter} as input
functions for describing the hadronization and convoluted them with
our previous results for the FFs of quarks and gluons (according to
the factorization theorem of QCD, see for example \cite{QCDrev}).
During the complete cascade, we paid a special attention to the
conservation of energy (what was not doable in previous studies,
because of the incomplete treatment of the cascade). We are able to
follow the energy conservation on the complete evolution up to a few
per thousand.\newline

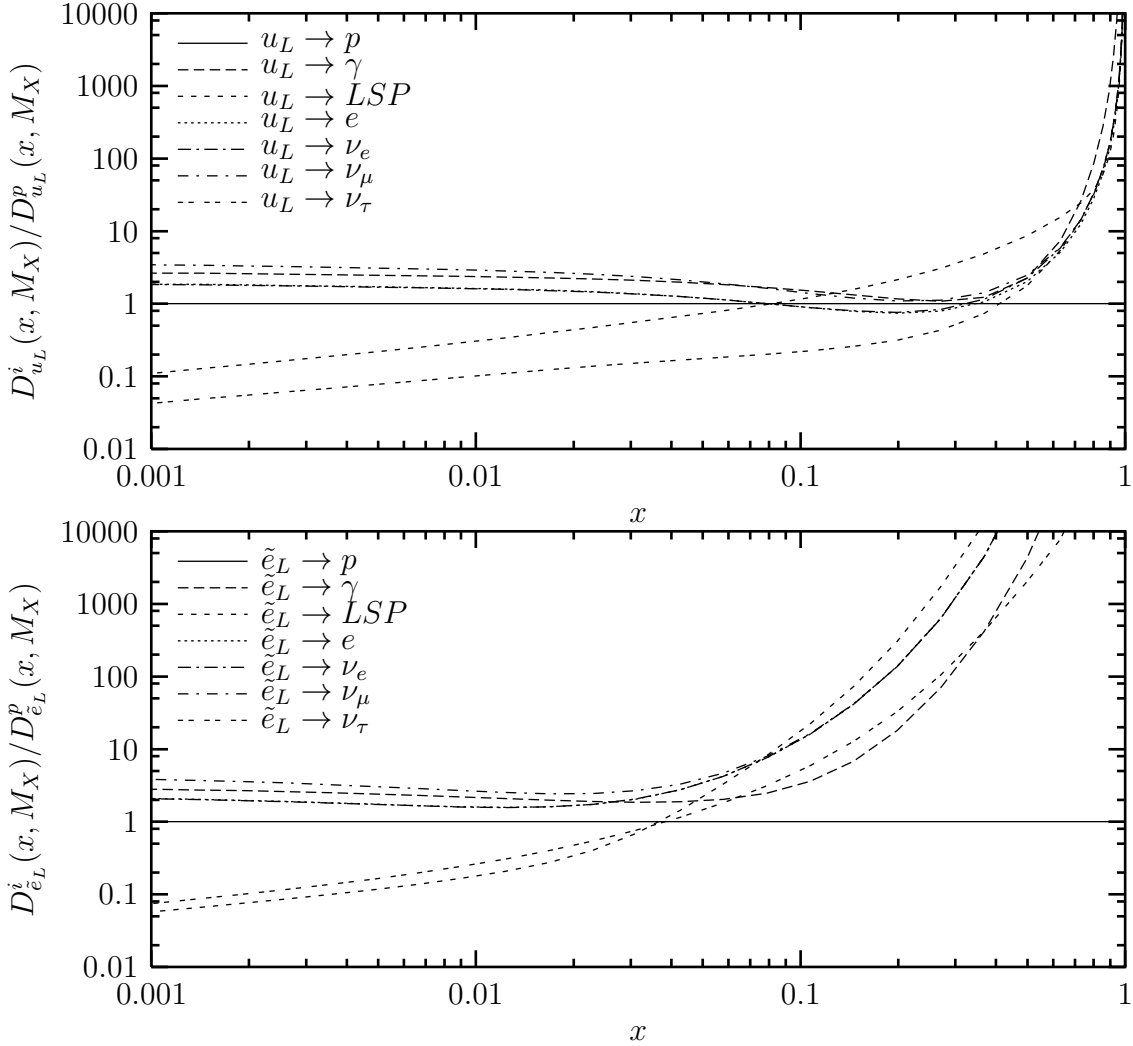
\begin{figure}[t!]
\setlength{\unitlength}{.3 cm}
\begin{minipage}[t]{4.5cm}
\input{Low_G_ratio_uL.tex}
\end{minipage}

\noindent
\begin{minipage}[b]{4.5cm}
\input{Low_G_ratio_eL_.tex}
\end{minipage} \hfill
\caption{Ratios of FFs $D_I^h/D_I^p$ for different stable particles $h$,
for an initial first or second generation $SU(2)$ doublet quark, $I =
q_L$, (top) or slepton, $I = \tilde e_L$, (bottom).} 
\label{ratio}
\end{figure}

As a result of the code we have written, we can obtain the spectrum of
any stable particle (protons, photons, electrons, neutrinos,
LSPs)\footnote{We summed over particles and antiparticles.} at the end
of the cascade, for {\it any} given N-body decay mode of the X
particle into particles of the MSSM\footnote{To describe any N body
  decay from the FFs of the decay products of X, we just need one more
  convolution between these FFs and the phase space of the decay.}. We
choose to show here the results obtained for two initial SU(2)
doublets of the first generation, a quark and a slepton.
Fig~\ref{FFs} shows the FFs themselves and fig~\ref{ratio} the ratios
of the different FFs on the proton flux.  Among the new general
features appearing on these spectra, I would like to describe here a
few points of particular interest :
\begin{itemize}
\item[1.] The spectra at small and large x are quite different ;
  especially, as can be seen on fig~\ref{ratio}, the small x physics
  is to a strong extent independant on the nature of the primaries,
  and the ratios between the different stable particles always remain
  the same. It allows us to order the fluxes from the strongest to the
  lowest one in the small x region : $\nu_\mu$, $\gamma$, $\nu_e$,
  electrons, and protons, within one order of magnitude; the two
  smallest fluxes, LSP and finally $\nu_\tau$, can be one order of
  magnitude lower than the proton flux at small x. We see that at
  large x, the fluxes are much more model dependent and change with
  the nature of the X decay products, but that LSP and $\gamma$ are
  generally the strongest ones.
\item[2.]  Due to the strength of the electroweak couplings at very
  high energy, the photon and neutrino spectra are even stronger than
  what was expected before ; these primaries have to be added to the
  flux of secondaries expected from propagation effects (especially
  through decay of pions after pion photo-production over the CMB). It
  is a puzzling result, because there are already strong indications
  that no photon event has been observed in the extremity of the CR
  spectrum (see for example \cite{noGamma:1}).\newline
\end{itemize}

In this talk, I have shown the type of results that can be obtained
with the code ``SHdecay'', which allows to compute the final spectra
of stable particles for any N-body decay mode of an initial
super-heavy X particle, in the framework of the MSSM. This code will
be soon made available. Combined with a model describing the nature,
the cosmic distribution, and the decaying properties of an X particle,
on one hand, and with a propagation code describing the losses of
energy of the stable particles traveling through the interstellar medium,
on the other hand, it allows to study many particular scenarios and to
make quantitative predictions for the fluxes to be observed on Earth
for each of these scenarios. We already developped a partial approach
of these issues, in collaboration with F. Halzen and D. Hooper, for
computing the expected fluxes of neutrinos \cite{bdhh:1} and
neutralinos \cite{bdhh:2} on Earth for different models, in the next
generation of experiments.

%--------------------------- References -------------------------------

\bibliographystyle{utphys}
\bibliography{references}

\clearpage
\noindent

\end{document}

%% file: Low_G_uL.tex
% GNUPLOT: LaTeX picture with Postscript
\begingroup%
  \makeatletter%
  \newcommand{\GNUPLOTspecial}{%
    \@sanitize\catcode`\%=14\relax\special}%
  \setlength{\unitlength}{0.1bp}%
{\GNUPLOTspecial{!
%!PS-Adobe-2.0 EPSF-2.0
%%Title: /a/srv-1.cip/h/t30i/dre/barbot/Latex/Talk_Susy02/Low_G_uL.tex
%%Creator: gnuplot 3.7 patchlevel 1
%%CreationDate: Thu Oct 17 15:32:04 2002
%%DocumentFonts: 
%%BoundingBox: 0 0 432 172
%%Orientation: Landscape
%%EndComments
/gnudict 256 dict def
gnudict begin
/Color false def
/Solid false def
/gnulinewidth 5.000 def
/userlinewidth gnulinewidth def
/vshift -33 def
/dl {10 mul} def
/hpt_ 31.5 def
/vpt_ 31.5 def
/hpt hpt_ def
/vpt vpt_ def
/M {moveto} bind def
/L {lineto} bind def
/R {rmoveto} bind def
/V {rlineto} bind def
/vpt2 vpt 2 mul def
/hpt2 hpt 2 mul def
/Lshow { currentpoint stroke M
  0 vshift R show } def
/Rshow { currentpoint stroke M
  dup stringwidth pop neg vshift R show } def
/Cshow { currentpoint stroke M
  dup stringwidth pop -2 div vshift R show } def
/UP { dup vpt_ mul /vpt exch def hpt_ mul /hpt exch def
  /hpt2 hpt 2 mul def /vpt2 vpt 2 mul def } def
/DL { Color {setrgbcolor Solid {pop []} if 0 setdash }
 {pop pop pop Solid {pop []} if 0 setdash} ifelse } def
/BL { stroke userlinewidth 2 mul setlinewidth } def
/AL { stroke userlinewidth 2 div setlinewidth } def
/UL { dup gnulinewidth mul /userlinewidth exch def
      10 mul /udl exch def } def
/PL { stroke userlinewidth setlinewidth } def
/LTb { BL [] 0 0 0 DL } def
/LTa { AL [1 udl mul 2 udl mul] 0 setdash 0 0 0 setrgbcolor } def
/LT0 { PL [] 1 0 0 DL } def
/LT1 { PL [4 dl 2 dl] 0 1 0 DL } def
/LT2 { PL [2 dl 3 dl] 0 0 1 DL } def
/LT3 { PL [1 dl 1.5 dl] 1 0 1 DL } def
/LT4 { PL [5 dl 2 dl 1 dl 2 dl] 0 1 1 DL } def
/LT5 { PL [4 dl 3 dl 1 dl 3 dl] 1 1 0 DL } def
/LT6 { PL [2 dl 2 dl 2 dl 4 dl] 0 0 0 DL } def
/LT7 { PL [2 dl 2 dl 2 dl 2 dl 2 dl 4 dl] 1 0.3 0 DL } def
/LT8 { PL [2 dl 2 dl 2 dl 2 dl 2 dl 2 dl 2 dl 4 dl] 0.5 0.5 0.5 DL } def
/Pnt { stroke [] 0 setdash
   gsave 1 setlinecap M 0 0 V stroke grestore } def
/Dia { stroke [] 0 setdash 2 copy vpt add M
  hpt neg vpt neg V hpt vpt neg V
  hpt vpt V hpt neg vpt V closepath stroke
  Pnt } def
/Pls { stroke [] 0 setdash vpt sub M 0 vpt2 V
  currentpoint stroke M
  hpt neg vpt neg R hpt2 0 V stroke
  } def
/Box { stroke [] 0 setdash 2 copy exch hpt sub exch vpt add M
  0 vpt2 neg V hpt2 0 V 0 vpt2 V
  hpt2 neg 0 V closepath stroke
  Pnt } def
/Crs { stroke [] 0 setdash exch hpt sub exch vpt add M
  hpt2 vpt2 neg V currentpoint stroke M
  hpt2 neg 0 R hpt2 vpt2 V stroke } def
/TriU { stroke [] 0 setdash 2 copy vpt 1.12 mul add M
  hpt neg vpt -1.62 mul V
  hpt 2 mul 0 V
  hpt neg vpt 1.62 mul V closepath stroke
  Pnt  } def
/Star { 2 copy Pls Crs } def
/BoxF { stroke [] 0 setdash exch hpt sub exch vpt add M
  0 vpt2 neg V  hpt2 0 V  0 vpt2 V
  hpt2 neg 0 V  closepath fill } def
/TriUF { stroke [] 0 setdash vpt 1.12 mul add M
  hpt neg vpt -1.62 mul V
  hpt 2 mul 0 V
  hpt neg vpt 1.62 mul V closepath fill } def
/TriD { stroke [] 0 setdash 2 copy vpt 1.12 mul sub M
  hpt neg vpt 1.62 mul V
  hpt 2 mul 0 V
  hpt neg vpt -1.62 mul V closepath stroke
  Pnt  } def
/TriDF { stroke [] 0 setdash vpt 1.12 mul sub M
  hpt neg vpt 1.62 mul V
  hpt 2 mul 0 V
  hpt neg vpt -1.62 mul V closepath fill} def
/DiaF { stroke [] 0 setdash vpt add M
  hpt neg vpt neg V hpt vpt neg V
  hpt vpt V hpt neg vpt V closepath fill } def
/Pent { stroke [] 0 setdash 2 copy gsave
  translate 0 hpt M 4 {72 rotate 0 hpt L} repeat
  closepath stroke grestore Pnt } def
/PentF { stroke [] 0 setdash gsave
  translate 0 hpt M 4 {72 rotate 0 hpt L} repeat
  closepath fill grestore } def
/Circle { stroke [] 0 setdash 2 copy
  hpt 0 360 arc stroke Pnt } def
/CircleF { stroke [] 0 setdash hpt 0 360 arc fill } def
/C0 { BL [] 0 setdash 2 copy moveto vpt 90 450  arc } bind def
/C1 { BL [] 0 setdash 2 copy        moveto
       2 copy  vpt 0 90 arc closepath fill
               vpt 0 360 arc closepath } bind def
/C2 { BL [] 0 setdash 2 copy moveto
       2 copy  vpt 90 180 arc closepath fill
               vpt 0 360 arc closepath } bind def
/C3 { BL [] 0 setdash 2 copy moveto
       2 copy  vpt 0 180 arc closepath fill
               vpt 0 360 arc closepath } bind def
/C4 { BL [] 0 setdash 2 copy moveto
       2 copy  vpt 180 270 arc closepath fill
               vpt 0 360 arc closepath } bind def
/C5 { BL [] 0 setdash 2 copy moveto
       2 copy  vpt 0 90 arc
       2 copy moveto
       2 copy  vpt 180 270 arc closepath fill
               vpt 0 360 arc } bind def
/C6 { BL [] 0 setdash 2 copy moveto
      2 copy  vpt 90 270 arc closepath fill
              vpt 0 360 arc closepath } bind def
/C7 { BL [] 0 setdash 2 copy moveto
      2 copy  vpt 0 270 arc closepath fill
              vpt 0 360 arc closepath } bind def
/C8 { BL [] 0 setdash 2 copy moveto
      2 copy vpt 270 360 arc closepath fill
              vpt 0 360 arc closepath } bind def
/C9 { BL [] 0 setdash 2 copy moveto
      2 copy  vpt 270 450 arc closepath fill
              vpt 0 360 arc closepath } bind def
/C10 { BL [] 0 setdash 2 copy 2 copy moveto vpt 270 360 arc closepath fill
       2 copy moveto
       2 copy vpt 90 180 arc closepath fill
               vpt 0 360 arc closepath } bind def
/C11 { BL [] 0 setdash 2 copy moveto
       2 copy  vpt 0 180 arc closepath fill
       2 copy moveto
       2 copy  vpt 270 360 arc closepath fill
               vpt 0 360 arc closepath } bind def
/C12 { BL [] 0 setdash 2 copy moveto
       2 copy  vpt 180 360 arc closepath fill
               vpt 0 360 arc closepath } bind def
/C13 { BL [] 0 setdash  2 copy moveto
       2 copy  vpt 0 90 arc closepath fill
       2 copy moveto
       2 copy  vpt 180 360 arc closepath fill
               vpt 0 360 arc closepath } bind def
/C14 { BL [] 0 setdash 2 copy moveto
       2 copy  vpt 90 360 arc closepath fill
               vpt 0 360 arc } bind def
/C15 { BL [] 0 setdash 2 copy vpt 0 360 arc closepath fill
               vpt 0 360 arc closepath } bind def
/Rec   { newpath 4 2 roll moveto 1 index 0 rlineto 0 exch rlineto
       neg 0 rlineto closepath } bind def
/Square { dup Rec } bind def
/Bsquare { vpt sub exch vpt sub exch vpt2 Square } bind def
/S0 { BL [] 0 setdash 2 copy moveto 0 vpt rlineto BL Bsquare } bind def
/S1 { BL [] 0 setdash 2 copy vpt Square fill Bsquare } bind def
/S2 { BL [] 0 setdash 2 copy exch vpt sub exch vpt Square fill Bsquare } bind def
/S3 { BL [] 0 setdash 2 copy exch vpt sub exch vpt2 vpt Rec fill Bsquare } bind def
/S4 { BL [] 0 setdash 2 copy exch vpt sub exch vpt sub vpt Square fill Bsquare } bind def
/S5 { BL [] 0 setdash 2 copy 2 copy vpt Square fill
       exch vpt sub exch vpt sub vpt Square fill Bsquare } bind def
/S6 { BL [] 0 setdash 2 copy exch vpt sub exch vpt sub vpt vpt2 Rec fill Bsquare } bind def
/S7 { BL [] 0 setdash 2 copy exch vpt sub exch vpt sub vpt vpt2 Rec fill
       2 copy vpt Square fill
       Bsquare } bind def
/S8 { BL [] 0 setdash 2 copy vpt sub vpt Square fill Bsquare } bind def
/S9 { BL [] 0 setdash 2 copy vpt sub vpt vpt2 Rec fill Bsquare } bind def
/S10 { BL [] 0 setdash 2 copy vpt sub vpt Square fill 2 copy exch vpt sub exch vpt Square fill
       Bsquare } bind def
/S11 { BL [] 0 setdash 2 copy vpt sub vpt Square fill 2 copy exch vpt sub exch vpt2 vpt Rec fill
       Bsquare } bind def
/S12 { BL [] 0 setdash 2 copy exch vpt sub exch vpt sub vpt2 vpt Rec fill Bsquare } bind def
/S13 { BL [] 0 setdash 2 copy exch vpt sub exch vpt sub vpt2 vpt Rec fill
       2 copy vpt Square fill Bsquare } bind def
/S14 { BL [] 0 setdash 2 copy exch vpt sub exch vpt sub vpt2 vpt Rec fill
       2 copy exch vpt sub exch vpt Square fill Bsquare } bind def
/S15 { BL [] 0 setdash 2 copy Bsquare fill Bsquare } bind def
/D0 { gsave translate 45 rotate 0 0 S0 stroke grestore } bind def
/D1 { gsave translate 45 rotate 0 0 S1 stroke grestore } bind def
/D2 { gsave translate 45 rotate 0 0 S2 stroke grestore } bind def
/D3 { gsave translate 45 rotate 0 0 S3 stroke grestore } bind def
/D4 { gsave translate 45 rotate 0 0 S4 stroke grestore } bind def
/D5 { gsave translate 45 rotate 0 0 S5 stroke grestore } bind def
/D6 { gsave translate 45 rotate 0 0 S6 stroke grestore } bind def
/D7 { gsave translate 45 rotate 0 0 S7 stroke grestore } bind def
/D8 { gsave translate 45 rotate 0 0 S8 stroke grestore } bind def
/D9 { gsave translate 45 rotate 0 0 S9 stroke grestore } bind def
/D10 { gsave translate 45 rotate 0 0 S10 stroke grestore } bind def
/D11 { gsave translate 45 rotate 0 0 S11 stroke grestore } bind def
/D12 { gsave translate 45 rotate 0 0 S12 stroke grestore } bind def
/D13 { gsave translate 45 rotate 0 0 S13 stroke grestore } bind def
/D14 { gsave translate 45 rotate 0 0 S14 stroke grestore } bind def
/D15 { gsave translate 45 rotate 0 0 S15 stroke grestore } bind def
/DiaE { stroke [] 0 setdash vpt add M
  hpt neg vpt neg V hpt vpt neg V
  hpt vpt V hpt neg vpt V closepath stroke } def
/BoxE { stroke [] 0 setdash exch hpt sub exch vpt add M
  0 vpt2 neg V hpt2 0 V 0 vpt2 V
  hpt2 neg 0 V closepath stroke } def
/TriUE { stroke [] 0 setdash vpt 1.12 mul add M
  hpt neg vpt -1.62 mul V
  hpt 2 mul 0 V
  hpt neg vpt 1.62 mul V closepath stroke } def
/TriDE { stroke [] 0 setdash vpt 1.12 mul sub M
  hpt neg vpt 1.62 mul V
  hpt 2 mul 0 V
  hpt neg vpt -1.62 mul V closepath stroke } def
/PentE { stroke [] 0 setdash gsave
  translate 0 hpt M 4 {72 rotate 0 hpt L} repeat
  closepath stroke grestore } def
/CircE { stroke [] 0 setdash 
  hpt 0 360 arc stroke } def
/Opaque { gsave closepath 1 setgray fill grestore 0 setgray closepath } def
/DiaW { stroke [] 0 setdash vpt add M
  hpt neg vpt neg V hpt vpt neg V
  hpt vpt V hpt neg vpt V Opaque stroke } def
/BoxW { stroke [] 0 setdash exch hpt sub exch vpt add M
  0 vpt2 neg V hpt2 0 V 0 vpt2 V
  hpt2 neg 0 V Opaque stroke } def
/TriUW { stroke [] 0 setdash vpt 1.12 mul add M
  hpt neg vpt -1.62 mul V
  hpt 2 mul 0 V
  hpt neg vpt 1.62 mul V Opaque stroke } def
/TriDW { stroke [] 0 setdash vpt 1.12 mul sub M
  hpt neg vpt 1.62 mul V
  hpt 2 mul 0 V
  hpt neg vpt -1.62 mul V Opaque stroke } def
/PentW { stroke [] 0 setdash gsave
  translate 0 hpt M 4 {72 rotate 0 hpt L} repeat
  Opaque stroke grestore } def
/CircW { stroke [] 0 setdash 
  hpt 0 360 arc Opaque stroke } def
/BoxFill { gsave Rec 1 setgray fill grestore } def
/Symbol-Oblique /Symbol findfont [1 0 .167 1 0 0] makefont
dup length dict begin {1 index /FID eq {pop pop} {def} ifelse} forall
currentdict end definefont
end
}}%
\begin{picture}(4320,1728)(0,0)%
{\GNUPLOTspecial{"
gnudict begin
gsave
0 0 translate
0.100 0.100 scale
0 setgray
newpath
1.000 UL
LTb
550 300 M
63 0 V
3557 0 R
-63 0 V
550 380 M
31 0 V
3589 0 R
-31 0 V
550 486 M
31 0 V
3589 0 R
-31 0 V
550 540 M
31 0 V
3589 0 R
-31 0 V
550 566 M
63 0 V
3557 0 R
-63 0 V
550 646 M
31 0 V
3589 0 R
-31 0 V
550 751 M
31 0 V
3589 0 R
-31 0 V
550 805 M
31 0 V
3589 0 R
-31 0 V
550 831 M
63 0 V
3557 0 R
-63 0 V
550 911 M
31 0 V
3589 0 R
-31 0 V
550 1017 M
31 0 V
3589 0 R
-31 0 V
550 1071 M
31 0 V
3589 0 R
-31 0 V
550 1097 M
63 0 V
3557 0 R
-63 0 V
550 1177 M
31 0 V
3589 0 R
-31 0 V
550 1282 M
31 0 V
3589 0 R
-31 0 V
550 1337 M
31 0 V
3589 0 R
-31 0 V
550 1362 M
63 0 V
3557 0 R
-63 0 V
550 1442 M
31 0 V
3589 0 R
-31 0 V
550 1548 M
31 0 V
3589 0 R
-31 0 V
550 1602 M
31 0 V
3589 0 R
-31 0 V
550 1628 M
63 0 V
3557 0 R
-63 0 V
550 300 M
0 63 V
0 1265 R
0 -63 V
913 300 M
0 31 V
0 1297 R
0 -31 V
1126 300 M
0 31 V
0 1297 R
0 -31 V
1276 300 M
0 31 V
0 1297 R
0 -31 V
1393 300 M
0 31 V
0 1297 R
0 -31 V
1489 300 M
0 31 V
0 1297 R
0 -31 V
1570 300 M
0 31 V
0 1297 R
0 -31 V
1640 300 M
0 31 V
0 1297 R
0 -31 V
1701 300 M
0 31 V
0 1297 R
0 -31 V
1757 300 M
0 63 V
0 1265 R
0 -63 V
2120 300 M
0 31 V
0 1297 R
0 -31 V
2332 300 M
0 31 V
0 1297 R
0 -31 V
2483 300 M
0 31 V
0 1297 R
0 -31 V
2600 300 M
0 31 V
0 1297 R
0 -31 V
2696 300 M
0 31 V
0 1297 R
0 -31 V
2776 300 M
0 31 V
0 1297 R
0 -31 V
2846 300 M
0 31 V
0 1297 R
0 -31 V
2908 300 M
0 31 V
0 1297 R
0 -31 V
2963 300 M
0 63 V
0 1265 R
0 -63 V
3327 300 M
0 31 V
0 1297 R
0 -31 V
3539 300 M
0 31 V
0 1297 R
0 -31 V
3690 300 M
0 31 V
0 1297 R
0 -31 V
3807 300 M
0 31 V
0 1297 R
0 -31 V
3902 300 M
0 31 V
0 1297 R
0 -31 V
3983 300 M
0 31 V
0 1297 R
0 -31 V
4053 300 M
0 31 V
0 1297 R
0 -31 V
4115 300 M
0 31 V
0 1297 R
0 -31 V
4170 300 M
0 63 V
0 1265 R
0 -63 V
1.000 UL
LTb
550 300 M
3620 0 V
0 1328 V
-3620 0 V
550 300 L
1.000 UL
LT0
650 1515 M
263 0 V
4056 300 M
-2 7 V
-49 139 V
-76 144 V
3806 740 L
3646 842 L
-161 50 V
-162 14 V
-162 -5 V
3000 882 L
2838 856 L
2676 824 L
2514 789 L
2353 752 L
2191 714 L
2029 675 L
1868 635 L
1706 594 L
1544 553 L
1383 511 L
1221 469 L
1059 426 L
898 382 L
736 337 L
597 300 L
1.000 UL
LT1
650 1415 M
263 0 V
4170 300 M
0 19 V
0 38 V
0 39 V
0 39 V
0 38 V
0 39 V
0 38 V
0 36 V
0 35 V
0 32 V
-1 25 V
0 19 V
0 12 V
0 9 V
-1 7 V
-1 8 V
-1 8 V
-1 8 V
-2 8 V
-2 8 V
-3 9 V
-5 9 V
-6 8 V
-8 9 V
-12 7 V
-16 7 V
-24 5 V
-33 2 V
-49 0 V
-76 0 V
-123 12 V
-160 33 V
-161 35 V
-162 24 V
-162 8 V
-161 -4 V
2838 915 L
2676 894 L
2514 868 L
2353 837 L
2191 804 L
2029 769 L
1868 732 L
1706 694 L
1544 656 L
1383 616 L
1221 575 L
1059 534 L
898 492 L
736 449 L
574 406 L
-24 -7 V
1.000 UL
LT2
650 1315 M
263 0 V
4160 300 M
-2 26 V
-5 46 V
-6 47 V
-8 48 V
-12 52 V
-16 58 V
-24 68 V
-33 79 V
-49 90 V
-76 93 V
-123 81 V
-160 37 V
-161 -4 V
3323 994 L
3161 953 L
3000 905 L
2838 853 L
2676 799 L
2514 742 L
2353 685 L
2191 627 L
2029 569 L
1868 511 L
1706 452 L
1544 394 L
1383 335 L
-98 -35 V
1.000 UL
LT3
650 1215 M
263 0 V
4159 300 M
-1 28 V
-5 56 V
-6 55 V
-8 54 V
-12 54 V
-16 51 V
-24 49 V
-33 46 V
-49 44 V
-76 42 V
-123 41 V
-160 30 V
-161 15 V
-162 7 V
-162 1 V
-161 -5 V
2838 856 L
2676 839 L
2514 816 L
2353 789 L
2191 758 L
2029 724 L
1868 688 L
1706 651 L
1544 613 L
1383 573 L
1221 533 L
1059 492 L
898 450 L
736 408 L
574 365 L
-24 -6 V
1.000 UL
LT4
650 1115 M
263 0 V
4160 300 M
-2 40 V
-5 57 V
-6 56 V
-8 55 V
-12 54 V
-16 52 V
-24 50 V
-33 46 V
-49 44 V
-76 41 V
-123 38 V
-160 26 V
-161 12 V
-162 4 V
-162 0 V
-161 -6 V
2838 856 L
2676 838 L
2514 815 L
2353 788 L
2191 757 L
2029 723 L
1868 687 L
1706 650 L
1544 612 L
1383 572 L
1221 532 L
1059 491 L
898 449 L
736 407 L
574 364 L
-24 -7 V
1.000 UL
LT5
650 1015 M
263 0 V
4160 300 M
-2 37 V
-5 57 V
-6 56 V
-8 55 V
-12 53 V
-16 52 V
-24 50 V
-33 47 V
-49 45 V
-76 45 V
-123 47 V
-160 38 V
-161 23 V
-162 12 V
-162 4 V
-161 -1 V
-162 -8 V
2676 897 L
2514 877 L
2353 852 L
2191 822 L
2029 790 L
1868 755 L
1706 718 L
1544 681 L
1383 642 L
1221 602 L
1059 561 L
898 520 L
736 478 L
574 436 L
-24 -7 V
1.000 UL
LT6
650 915 M
263 0 V
4160 300 M
-2 40 V
-5 57 V
-6 57 V
-8 56 V
-12 54 V
-16 52 V
-24 50 V
-33 46 V
-49 41 V
-76 32 V
-123 20 V
-160 2 V
3485 795 L
3323 773 L
3161 745 L
3000 710 L
2838 671 L
2676 629 L
2514 583 L
2353 535 L
2191 485 L
2029 433 L
1868 380 L
1706 327 L
-78 -27 V
stroke
grestore
end
showpage
}}%
\put(963,915){\makebox(0,0)[l]{$u_L \rightarrow \nu_\tau$}}%
\put(963,1015){\makebox(0,0)[l]{$u_L \rightarrow \nu_\mu$}}%
\put(963,1115){\makebox(0,0)[l]{$u_L \rightarrow \nu_e$}}%
\put(963,1215){\makebox(0,0)[l]{$u_L \rightarrow e$}}%
\put(963,1315){\makebox(0,0)[l]{$u_L \rightarrow LSP$}}%
\put(963,1415){\makebox(0,0)[l]{$u_L \rightarrow \gamma$}}%
\put(963,1515){\makebox(0,0)[l]{$u_L \rightarrow p$}}%
\put(2360,50){\makebox(0,0){$x$}}%
\put(100,964){%
\special{ps: gsave currentpoint currentpoint translate
270 rotate neg exch neg exch translate}%
\makebox(0,0)[b]{\shortstack{$x^3 D_{u_L}^i(x,M_X)$}}%
\special{ps: currentpoint grestore moveto}%
}%
\put(4170,200){\makebox(0,0){1}}%
\put(2963,200){\makebox(0,0){0.1}}%
\put(1757,200){\makebox(0,0){0.01}}%
\put(550,200){\makebox(0,0){0.001}}%
\put(500,1628){\makebox(0,0)[r]{1}}%
\put(500,1362){\makebox(0,0)[r]{0.1}}%
\put(500,1097){\makebox(0,0)[r]{0.01}}%
\put(500,831){\makebox(0,0)[r]{0.001}}%
\put(500,566){\makebox(0,0)[r]{0.0001}}%
\put(500,300){\makebox(0,0)[r]{1e-05}}%
\end{picture}%
\endgroup
 

%% file: Low_G_eL_.tex
% GNUPLOT: LaTeX picture with Postscript
\begingroup%
  \makeatletter%
  \newcommand{\GNUPLOTspecial}{%
    \@sanitize\catcode`\%=14\relax\special}%
  \setlength{\unitlength}{0.1bp}%
{\GNUPLOTspecial{!
%!PS-Adobe-2.0 EPSF-2.0
%%Title: /a/srv-1.cip/h/t30i/dre/barbot/Latex/Talk_Susy02/Low_G_eL_.tex
%%Creator: gnuplot 3.7 patchlevel 1
%%CreationDate: Thu Oct 17 15:32:05 2002
%%DocumentFonts: 
%%BoundingBox: 0 0 432 172
%%Orientation: Landscape
%%EndComments
/gnudict 256 dict def
gnudict begin
/Color false def
/Solid false def
/gnulinewidth 5.000 def
/userlinewidth gnulinewidth def
/vshift -33 def
/dl {10 mul} def
/hpt_ 31.5 def
/vpt_ 31.5 def
/hpt hpt_ def
/vpt vpt_ def
/M {moveto} bind def
/L {lineto} bind def
/R {rmoveto} bind def
/V {rlineto} bind def
/vpt2 vpt 2 mul def
/hpt2 hpt 2 mul def
/Lshow { currentpoint stroke M
  0 vshift R show } def
/Rshow { currentpoint stroke M
  dup stringwidth pop neg vshift R show } def
/Cshow { currentpoint stroke M
  dup stringwidth pop -2 div vshift R show } def
/UP { dup vpt_ mul /vpt exch def hpt_ mul /hpt exch def
  /hpt2 hpt 2 mul def /vpt2 vpt 2 mul def } def
/DL { Color {setrgbcolor Solid {pop []} if 0 setdash }
 {pop pop pop Solid {pop []} if 0 setdash} ifelse } def
/BL { stroke userlinewidth 2 mul setlinewidth } def
/AL { stroke userlinewidth 2 div setlinewidth } def
/UL { dup gnulinewidth mul /userlinewidth exch def
      10 mul /udl exch def } def
/PL { stroke userlinewidth setlinewidth } def
/LTb { BL [] 0 0 0 DL } def
/LTa { AL [1 udl mul 2 udl mul] 0 setdash 0 0 0 setrgbcolor } def
/LT0 { PL [] 1 0 0 DL } def
/LT1 { PL [4 dl 2 dl] 0 1 0 DL } def
/LT2 { PL [2 dl 3 dl] 0 0 1 DL } def
/LT3 { PL [1 dl 1.5 dl] 1 0 1 DL } def
/LT4 { PL [5 dl 2 dl 1 dl 2 dl] 0 1 1 DL } def
/LT5 { PL [4 dl 3 dl 1 dl 3 dl] 1 1 0 DL } def
/LT6 { PL [2 dl 2 dl 2 dl 4 dl] 0 0 0 DL } def
/LT7 { PL [2 dl 2 dl 2 dl 2 dl 2 dl 4 dl] 1 0.3 0 DL } def
/LT8 { PL [2 dl 2 dl 2 dl 2 dl 2 dl 2 dl 2 dl 4 dl] 0.5 0.5 0.5 DL } def
/Pnt { stroke [] 0 setdash
   gsave 1 setlinecap M 0 0 V stroke grestore } def
/Dia { stroke [] 0 setdash 2 copy vpt add M
  hpt neg vpt neg V hpt vpt neg V
  hpt vpt V hpt neg vpt V closepath stroke
  Pnt } def
/Pls { stroke [] 0 setdash vpt sub M 0 vpt2 V
  currentpoint stroke M
  hpt neg vpt neg R hpt2 0 V stroke
  } def
/Box { stroke [] 0 setdash 2 copy exch hpt sub exch vpt add M
  0 vpt2 neg V hpt2 0 V 0 vpt2 V
  hpt2 neg 0 V closepath stroke
  Pnt } def
/Crs { stroke [] 0 setdash exch hpt sub exch vpt add M
  hpt2 vpt2 neg V currentpoint stroke M
  hpt2 neg 0 R hpt2 vpt2 V stroke } def
/TriU { stroke [] 0 setdash 2 copy vpt 1.12 mul add M
  hpt neg vpt -1.62 mul V
  hpt 2 mul 0 V
  hpt neg vpt 1.62 mul V closepath stroke
  Pnt  } def
/Star { 2 copy Pls Crs } def
/BoxF { stroke [] 0 setdash exch hpt sub exch vpt add M
  0 vpt2 neg V  hpt2 0 V  0 vpt2 V
  hpt2 neg 0 V  closepath fill } def
/TriUF { stroke [] 0 setdash vpt 1.12 mul add M
  hpt neg vpt -1.62 mul V
  hpt 2 mul 0 V
  hpt neg vpt 1.62 mul V closepath fill } def
/TriD { stroke [] 0 setdash 2 copy vpt 1.12 mul sub M
  hpt neg vpt 1.62 mul V
  hpt 2 mul 0 V
  hpt neg vpt -1.62 mul V closepath stroke
  Pnt  } def
/TriDF { stroke [] 0 setdash vpt 1.12 mul sub M
  hpt neg vpt 1.62 mul V
  hpt 2 mul 0 V
  hpt neg vpt -1.62 mul V closepath fill} def
/DiaF { stroke [] 0 setdash vpt add M
  hpt neg vpt neg V hpt vpt neg V
  hpt vpt V hpt neg vpt V closepath fill } def
/Pent { stroke [] 0 setdash 2 copy gsave
  translate 0 hpt M 4 {72 rotate 0 hpt L} repeat
  closepath stroke grestore Pnt } def
/PentF { stroke [] 0 setdash gsave
  translate 0 hpt M 4 {72 rotate 0 hpt L} repeat
  closepath fill grestore } def
/Circle { stroke [] 0 setdash 2 copy
  hpt 0 360 arc stroke Pnt } def
/CircleF { stroke [] 0 setdash hpt 0 360 arc fill } def
/C0 { BL [] 0 setdash 2 copy moveto vpt 90 450  arc } bind def
/C1 { BL [] 0 setdash 2 copy        moveto
       2 copy  vpt 0 90 arc closepath fill
               vpt 0 360 arc closepath } bind def
/C2 { BL [] 0 setdash 2 copy moveto
       2 copy  vpt 90 180 arc closepath fill
               vpt 0 360 arc closepath } bind def
/C3 { BL [] 0 setdash 2 copy moveto
       2 copy  vpt 0 180 arc closepath fill
               vpt 0 360 arc closepath } bind def
/C4 { BL [] 0 setdash 2 copy moveto
       2 copy  vpt 180 270 arc closepath fill
               vpt 0 360 arc closepath } bind def
/C5 { BL [] 0 setdash 2 copy moveto
       2 copy  vpt 0 90 arc
       2 copy moveto
       2 copy  vpt 180 270 arc closepath fill
               vpt 0 360 arc } bind def
/C6 { BL [] 0 setdash 2 copy moveto
      2 copy  vpt 90 270 arc closepath fill
              vpt 0 360 arc closepath } bind def
/C7 { BL [] 0 setdash 2 copy moveto
      2 copy  vpt 0 270 arc closepath fill
              vpt 0 360 arc closepath } bind def
/C8 { BL [] 0 setdash 2 copy moveto
      2 copy vpt 270 360 arc closepath fill
              vpt 0 360 arc closepath } bind def
/C9 { BL [] 0 setdash 2 copy moveto
      2 copy  vpt 270 450 arc closepath fill
              vpt 0 360 arc closepath } bind def
/C10 { BL [] 0 setdash 2 copy 2 copy moveto vpt 270 360 arc closepath fill
       2 copy moveto
       2 copy vpt 90 180 arc closepath fill
               vpt 0 360 arc closepath } bind def
/C11 { BL [] 0 setdash 2 copy moveto
       2 copy  vpt 0 180 arc closepath fill
       2 copy moveto
       2 copy  vpt 270 360 arc closepath fill
               vpt 0 360 arc closepath } bind def
/C12 { BL [] 0 setdash 2 copy moveto
       2 copy  vpt 180 360 arc closepath fill
               vpt 0 360 arc closepath } bind def
/C13 { BL [] 0 setdash  2 copy moveto
       2 copy  vpt 0 90 arc closepath fill
       2 copy moveto
       2 copy  vpt 180 360 arc closepath fill
               vpt 0 360 arc closepath } bind def
/C14 { BL [] 0 setdash 2 copy moveto
       2 copy  vpt 90 360 arc closepath fill
               vpt 0 360 arc } bind def
/C15 { BL [] 0 setdash 2 copy vpt 0 360 arc closepath fill
               vpt 0 360 arc closepath } bind def
/Rec   { newpath 4 2 roll moveto 1 index 0 rlineto 0 exch rlineto
       neg 0 rlineto closepath } bind def
/Square { dup Rec } bind def
/Bsquare { vpt sub exch vpt sub exch vpt2 Square } bind def
/S0 { BL [] 0 setdash 2 copy moveto 0 vpt rlineto BL Bsquare } bind def
/S1 { BL [] 0 setdash 2 copy vpt Square fill Bsquare } bind def
/S2 { BL [] 0 setdash 2 copy exch vpt sub exch vpt Square fill Bsquare } bind def
/S3 { BL [] 0 setdash 2 copy exch vpt sub exch vpt2 vpt Rec fill Bsquare } bind def
/S4 { BL [] 0 setdash 2 copy exch vpt sub exch vpt sub vpt Square fill Bsquare } bind def
/S5 { BL [] 0 setdash 2 copy 2 copy vpt Square fill
       exch vpt sub exch vpt sub vpt Square fill Bsquare } bind def
/S6 { BL [] 0 setdash 2 copy exch vpt sub exch vpt sub vpt vpt2 Rec fill Bsquare } bind def
/S7 { BL [] 0 setdash 2 copy exch vpt sub exch vpt sub vpt vpt2 Rec fill
       2 copy vpt Square fill
       Bsquare } bind def
/S8 { BL [] 0 setdash 2 copy vpt sub vpt Square fill Bsquare } bind def
/S9 { BL [] 0 setdash 2 copy vpt sub vpt vpt2 Rec fill Bsquare } bind def
/S10 { BL [] 0 setdash 2 copy vpt sub vpt Square fill 2 copy exch vpt sub exch vpt Square fill
       Bsquare } bind def
/S11 { BL [] 0 setdash 2 copy vpt sub vpt Square fill 2 copy exch vpt sub exch vpt2 vpt Rec fill
       Bsquare } bind def
/S12 { BL [] 0 setdash 2 copy exch vpt sub exch vpt sub vpt2 vpt Rec fill Bsquare } bind def
/S13 { BL [] 0 setdash 2 copy exch vpt sub exch vpt sub vpt2 vpt Rec fill
       2 copy vpt Square fill Bsquare } bind def
/S14 { BL [] 0 setdash 2 copy exch vpt sub exch vpt sub vpt2 vpt Rec fill
       2 copy exch vpt sub exch vpt Square fill Bsquare } bind def
/S15 { BL [] 0 setdash 2 copy Bsquare fill Bsquare } bind def
/D0 { gsave translate 45 rotate 0 0 S0 stroke grestore } bind def
/D1 { gsave translate 45 rotate 0 0 S1 stroke grestore } bind def
/D2 { gsave translate 45 rotate 0 0 S2 stroke grestore } bind def
/D3 { gsave translate 45 rotate 0 0 S3 stroke grestore } bind def
/D4 { gsave translate 45 rotate 0 0 S4 stroke grestore } bind def
/D5 { gsave translate 45 rotate 0 0 S5 stroke grestore } bind def
/D6 { gsave translate 45 rotate 0 0 S6 stroke grestore } bind def
/D7 { gsave translate 45 rotate 0 0 S7 stroke grestore } bind def
/D8 { gsave translate 45 rotate 0 0 S8 stroke grestore } bind def
/D9 { gsave translate 45 rotate 0 0 S9 stroke grestore } bind def
/D10 { gsave translate 45 rotate 0 0 S10 stroke grestore } bind def
/D11 { gsave translate 45 rotate 0 0 S11 stroke grestore } bind def
/D12 { gsave translate 45 rotate 0 0 S12 stroke grestore } bind def
/D13 { gsave translate 45 rotate 0 0 S13 stroke grestore } bind def
/D14 { gsave translate 45 rotate 0 0 S14 stroke grestore } bind def
/D15 { gsave translate 45 rotate 0 0 S15 stroke grestore } bind def
/DiaE { stroke [] 0 setdash vpt add M
  hpt neg vpt neg V hpt vpt neg V
  hpt vpt V hpt neg vpt V closepath stroke } def
/BoxE { stroke [] 0 setdash exch hpt sub exch vpt add M
  0 vpt2 neg V hpt2 0 V 0 vpt2 V
  hpt2 neg 0 V closepath stroke } def
/TriUE { stroke [] 0 setdash vpt 1.12 mul add M
  hpt neg vpt -1.62 mul V
  hpt 2 mul 0 V
  hpt neg vpt 1.62 mul V closepath stroke } def
/TriDE { stroke [] 0 setdash vpt 1.12 mul sub M
  hpt neg vpt 1.62 mul V
  hpt 2 mul 0 V
  hpt neg vpt -1.62 mul V closepath stroke } def
/PentE { stroke [] 0 setdash gsave
  translate 0 hpt M 4 {72 rotate 0 hpt L} repeat
  closepath stroke grestore } def
/CircE { stroke [] 0 setdash 
  hpt 0 360 arc stroke } def
/Opaque { gsave closepath 1 setgray fill grestore 0 setgray closepath } def
/DiaW { stroke [] 0 setdash vpt add M
  hpt neg vpt neg V hpt vpt neg V
  hpt vpt V hpt neg vpt V Opaque stroke } def
/BoxW { stroke [] 0 setdash exch hpt sub exch vpt add M
  0 vpt2 neg V hpt2 0 V 0 vpt2 V
  hpt2 neg 0 V Opaque stroke } def
/TriUW { stroke [] 0 setdash vpt 1.12 mul add M
  hpt neg vpt -1.62 mul V
  hpt 2 mul 0 V
  hpt neg vpt 1.62 mul V Opaque stroke } def
/TriDW { stroke [] 0 setdash vpt 1.12 mul sub M
  hpt neg vpt 1.62 mul V
  hpt 2 mul 0 V
  hpt neg vpt -1.62 mul V Opaque stroke } def
/PentW { stroke [] 0 setdash gsave
  translate 0 hpt M 4 {72 rotate 0 hpt L} repeat
  Opaque stroke grestore } def
/CircW { stroke [] 0 setdash 
  hpt 0 360 arc Opaque stroke } def
/BoxFill { gsave Rec 1 setgray fill grestore } def
/Symbol-Oblique /Symbol findfont [1 0 .167 1 0 0] makefont
dup length dict begin {1 index /FID eq {pop pop} {def} ifelse} forall
currentdict end definefont
end
}}%
\begin{picture}(4320,1728)(0,0)%
{\GNUPLOTspecial{"
gnudict begin
gsave
0 0 translate
0.100 0.100 scale
0 setgray
newpath
1.000 UL
LTb
550 300 M
63 0 V
3557 0 R
-63 0 V
550 380 M
31 0 V
3589 0 R
-31 0 V
550 486 M
31 0 V
3589 0 R
-31 0 V
550 540 M
31 0 V
3589 0 R
-31 0 V
550 566 M
63 0 V
3557 0 R
-63 0 V
550 646 M
31 0 V
3589 0 R
-31 0 V
550 751 M
31 0 V
3589 0 R
-31 0 V
550 805 M
31 0 V
3589 0 R
-31 0 V
550 831 M
63 0 V
3557 0 R
-63 0 V
550 911 M
31 0 V
3589 0 R
-31 0 V
550 1017 M
31 0 V
3589 0 R
-31 0 V
550 1071 M
31 0 V
3589 0 R
-31 0 V
550 1097 M
63 0 V
3557 0 R
-63 0 V
550 1177 M
31 0 V
3589 0 R
-31 0 V
550 1282 M
31 0 V
3589 0 R
-31 0 V
550 1337 M
31 0 V
3589 0 R
-31 0 V
550 1362 M
63 0 V
3557 0 R
-63 0 V
550 1442 M
31 0 V
3589 0 R
-31 0 V
550 1548 M
31 0 V
3589 0 R
-31 0 V
550 1602 M
31 0 V
3589 0 R
-31 0 V
550 1628 M
63 0 V
3557 0 R
-63 0 V
550 300 M
0 63 V
0 1265 R
0 -63 V
913 300 M
0 31 V
0 1297 R
0 -31 V
1126 300 M
0 31 V
0 1297 R
0 -31 V
1276 300 M
0 31 V
0 1297 R
0 -31 V
1393 300 M
0 31 V
0 1297 R
0 -31 V
1489 300 M
0 31 V
0 1297 R
0 -31 V
1570 300 M
0 31 V
0 1297 R
0 -31 V
1640 300 M
0 31 V
0 1297 R
0 -31 V
1701 300 M
0 31 V
0 1297 R
0 -31 V
1757 300 M
0 63 V
0 1265 R
0 -63 V
2120 300 M
0 31 V
0 1297 R
0 -31 V
2332 300 M
0 31 V
0 1297 R
0 -31 V
2483 300 M
0 31 V
0 1297 R
0 -31 V
2600 300 M
0 31 V
0 1297 R
0 -31 V
2696 300 M
0 31 V
0 1297 R
0 -31 V
2776 300 M
0 31 V
0 1297 R
0 -31 V
2846 300 M
0 31 V
0 1297 R
0 -31 V
2908 300 M
0 31 V
0 1297 R
0 -31 V
2963 300 M
0 63 V
0 1265 R
0 -63 V
3327 300 M
0 31 V
0 1297 R
0 -31 V
3539 300 M
0 31 V
0 1297 R
0 -31 V
3690 300 M
0 31 V
0 1297 R
0 -31 V
3807 300 M
0 31 V
0 1297 R
0 -31 V
3902 300 M
0 31 V
0 1297 R
0 -31 V
3983 300 M
0 31 V
0 1297 R
0 -31 V
4053 300 M
0 31 V
0 1297 R
0 -31 V
4115 300 M
0 31 V
0 1297 R
0 -31 V
4170 300 M
0 63 V
0 1265 R
0 -63 V
1.000 UL
LTb
550 300 M
3620 0 V
0 1328 V
-3620 0 V
550 300 L
1.000 UL
LT0
650 1506 M
263 0 V
3573 300 M
-88 87 V
3323 493 L
-162 68 V
-161 40 V
-162 19 V
-162 3 V
-162 -9 V
2353 595 L
2191 569 L
2029 537 L
1868 500 L
1706 460 L
1544 418 L
1383 373 L
1221 327 L
-95 -27 V
1.000 UL
LT1
650 1388 M
263 0 V
4169 300 M
0 6 V
0 56 V
0 52 V
-1 47 V
-1 45 V
-1 43 V
-1 42 V
-2 41 V
-2 40 V
-3 39 V
-5 39 V
-6 37 V
-8 37 V
-4 12 V
-2 8 V
-2 6 V
-2 4 V
-1 2 V
-1 2 V
0 2 V
-1 1 V
0 1 V
0 1 V
-1 0 V
0 1 V
0 1 V
-1 1 V
0 1 V
-1 1 V
-1 2 V
-1 2 V
-1 3 V
-2 4 V
-3 5 V
-4 7 V
-5 9 V
-8 10 V
-11 13 V
-33 26 V
-49 19 V
-76 8 V
3806 962 L
3646 925 L
3485 877 L
3323 827 L
3161 783 L
3000 749 L
2838 724 L
2676 705 L
2514 686 L
2353 666 L
2191 643 L
2029 616 L
1868 585 L
1706 551 L
1544 514 L
1383 475 L
1221 433 L
1059 390 L
898 346 L
736 301 L
-4 -1 V
1.000 UL
LT2
650 1270 M
263 0 V
4170 300 M
0 20 V
0 40 V
0 41 V
0 41 V
0 42 V
0 41 V
0 42 V
0 41 V
0 42 V
0 42 V
0 42 V
0 42 V
0 42 V
0 42 V
0 42 V
0 42 V
0 42 V
0 41 V
0 41 V
0 38 V
0 35 V
-1 30 V
0 22 V
0 16 V
0 11 V
-1 10 V
-1 9 V
-1 9 V
-1 8 V
-2 7 V
-2 8 V
-3 7 V
-5 6 V
-6 7 V
-8 6 V
-4 2 V
-2 2 V
-2 1 V
-2 0 V
-1 1 V
-1 0 V
0 1 V
-1 0 V
-1 1 V
-1 0 V
-1 1 V
-1 0 V
-1 1 V
-1 0 V
-2 1 V
-3 2 V
-4 2 V
-5 2 V
-8 4 V
-11 5 V
-33 15 V
-49 18 V
-76 14 V
-123 -6 V
-160 -44 V
-161 -74 V
-162 -93 V
-162 -97 V
3000 959 L
2838 862 L
2676 760 L
2514 655 L
2353 553 L
2191 465 L
2029 388 L
1868 320 L
-49 -20 V
1.000 UL
LT3
650 1152 M
263 0 V
4170 300 M
0 9 V
0 45 V
0 45 V
0 45 V
0 45 V
0 45 V
0 45 V
0 45 V
0 44 V
0 45 V
0 45 V
0 44 V
0 44 V
0 42 V
0 41 V
0 37 V
-1 32 V
0 25 V
0 18 V
0 14 V
-1 11 V
-1 11 V
-1 9 V
-1 10 V
-2 8 V
-2 9 V
-3 43 V
-5 20 V
-6 11 V
-8 7 V
-4 1 V
-2 1 V
-2 0 V
-2 1 V
-1 0 V
-1 0 V
-1 15 V
0 20 V
0 9 V
0 3 V
-1 8 V
0 10 V
0 8 V
0 5 V
0 4 V
0 2 V
0 2 V
0 2 V
0 1 V
0 1 V
0 1 V
0 1 V
0 1 V
0 1 V
0 1 V
0 2 V
0 2 V
0 2 V
0 3 V
0 4 V
0 4 V
-1 4 V
0 4 V
0 5 V
-1 5 V
-1 5 V
-1 6 V
-1 5 V
-2 6 V
-3 6 V
-4 5 V
-5 5 V
-8 3 V
-11 2 V
-33 -4 V
-49 -16 V
-76 -31 V
-123 -53 V
-160 -67 V
-161 -68 V
-162 -69 V
3161 992 L
3000 922 L
2838 853 L
2676 789 L
2514 730 L
2353 678 L
2191 632 L
2029 591 L
1868 553 L
1706 515 L
1544 477 L
1383 437 L
1221 397 L
1059 354 L
898 311 L
857 300 L
1.000 UL
LT4
650 1034 M
263 0 V
4170 300 M
0 9 V
0 45 V
0 45 V
0 45 V
0 45 V
0 45 V
0 45 V
0 45 V
0 44 V
0 45 V
0 45 V
0 44 V
0 44 V
0 42 V
0 41 V
0 37 V
-1 32 V
0 25 V
0 18 V
0 14 V
-1 11 V
-1 11 V
-1 9 V
-1 10 V
-2 8 V
-2 9 V
-3 46 V
-5 26 V
-6 11 V
-8 8 V
-4 1 V
-2 1 V
-2 1 V
-2 0 V
-1 0 V
-1 0 V
0 1 V
-1 17 V
0 35 V
0 14 V
0 6 V
-1 6 V
0 6 V
0 4 V
0 2 V
0 2 V
0 2 V
0 1 V
0 1 V
0 1 V
0 1 V
0 1 V
0 1 V
0 1 V
0 1 V
0 2 V
0 2 V
0 3 V
-1 2 V
0 4 V
0 3 V
-1 4 V
-1 5 V
-1 5 V
-1 5 V
-2 5 V
-3 6 V
-4 5 V
-5 4 V
-8 3 V
-11 2 V
-33 -5 V
-49 -16 V
-76 -30 V
-123 -53 V
-160 -67 V
-161 -69 V
-162 -69 V
3161 991 L
3000 922 L
2838 854 L
2676 789 L
2514 730 L
2353 678 L
2191 632 L
2029 591 L
1868 552 L
1706 514 L
1544 476 L
1383 436 L
1221 396 L
1059 353 L
898 310 L
861 300 L
1.000 UL
LT5
650 916 M
263 0 V
4170 300 M
0 9 V
0 45 V
0 45 V
0 45 V
0 45 V
0 45 V
0 45 V
0 45 V
0 44 V
0 45 V
0 45 V
0 44 V
0 44 V
0 42 V
0 41 V
0 37 V
-1 32 V
0 25 V
0 18 V
0 14 V
-1 11 V
-1 11 V
-1 9 V
-1 10 V
-2 8 V
-2 9 V
-3 46 V
-5 26 V
-6 11 V
-8 8 V
-4 1 V
-2 1 V
-2 1 V
-2 0 V
-1 0 V
-1 0 V
0 1 V
-1 17 V
0 35 V
0 14 V
0 6 V
-1 6 V
0 6 V
0 4 V
0 2 V
0 2 V
0 2 V
0 1 V
0 1 V
0 1 V
0 1 V
0 1 V
0 1 V
0 1 V
0 1 V
0 2 V
0 2 V
0 3 V
-1 2 V
0 4 V
0 3 V
-1 4 V
-1 5 V
-1 5 V
-1 5 V
-2 5 V
-3 6 V
-4 4 V
-5 5 V
-8 3 V
-11 2 V
-33 -5 V
-49 -16 V
-76 -30 V
-123 -53 V
-160 -68 V
-161 -68 V
-162 -69 V
3161 992 L
3000 924 L
2838 859 L
2676 801 L
2514 751 L
2353 708 L
2191 672 L
2029 639 L
1868 607 L
1706 574 L
1544 539 L
1383 502 L
1221 463 L
1059 422 L
898 380 L
736 336 L
607 300 L
1.000 UL
LT6
650 798 M
263 0 V
4139 300 M
0 1 V
-4 28 V
-2 18 V
-2 13 V
-2 9 V
-1 7 V
-1 5 V
0 3 V
-1 3 V
0 2 V
0 1 V
0 1 V
-1 1 V
0 1 V
0 1 V
0 1 V
0 1 V
0 1 V
-1 1 V
0 2 V
0 2 V
-1 3 V
-1 3 V
-1 6 V
-1 7 V
-2 9 V
-3 13 V
-4 17 V
-5 21 V
-8 27 V
-11 33 V
-33 74 V
-49 69 V
-76 85 V
3806 877 L
-160 49 V
-161 -2 V
3323 897 L
3161 856 L
3000 806 L
2838 751 L
2676 691 L
2514 629 L
2353 564 L
2191 498 L
2029 431 L
1868 364 L
1713 300 L
stroke
grestore
end
showpage
}}%
\put(963,798){\makebox(0,0)[l]{$\tilde{e}_L \rightarrow \nu_\tau$}}%
\put(963,916){\makebox(0,0)[l]{$\tilde{e}_L \rightarrow \nu_\mu$}}%
\put(963,1034){\makebox(0,0)[l]{$\tilde{e}_L \rightarrow \nu_e$}}%
\put(963,1152){\makebox(0,0)[l]{$\tilde{e}_L \rightarrow e$}}%
\put(963,1270){\makebox(0,0)[l]{$\tilde{e}_L \rightarrow LSP$}}%
\put(963,1388){\makebox(0,0)[l]{$\tilde{e}_L \rightarrow \gamma$}}%
\put(963,1506){\makebox(0,0)[l]{$\tilde{e}_L \rightarrow p$}}%
\put(2360,50){\makebox(0,0){$x$}}%
\put(100,964){%
\special{ps: gsave currentpoint currentpoint translate
270 rotate neg exch neg exch translate}%
\makebox(0,0)[b]{\shortstack{$x^3 D_{\tilde{e}_L}^i(x,M_X)$}}%
\special{ps: currentpoint grestore moveto}%
}%
\put(4170,200){\makebox(0,0){1}}%
\put(2963,200){\makebox(0,0){0.1}}%
\put(1757,200){\makebox(0,0){0.01}}%
\put(550,200){\makebox(0,0){0.001}}%
\put(500,1628){\makebox(0,0)[r]{1}}%
\put(500,1362){\makebox(0,0)[r]{0.1}}%
\put(500,1097){\makebox(0,0)[r]{0.01}}%
\put(500,831){\makebox(0,0)[r]{0.001}}%
\put(500,566){\makebox(0,0)[r]{0.0001}}%
\put(500,300){\makebox(0,0)[r]{1e-05}}%
\end{picture}%
\endgroup
 

%% file: Low_G_ratio_uL.tex
% GNUPLOT: LaTeX picture with Postscript
\begingroup%
  \makeatletter%
  \newcommand{\GNUPLOTspecial}{%
    \@sanitize\catcode`\%=14\relax\special}%
  \setlength{\unitlength}{0.1bp}%
{\GNUPLOTspecial{!
%!PS-Adobe-2.0 EPSF-2.0
%%Title: /a/srv-1.cip/h/t30i/dre/barbot/Latex/Talk_Susy02/Low_G_ratio_uL.tex
%%Creator: gnuplot 3.7 patchlevel 1
%%CreationDate: Thu Oct 17 15:38:18 2002
%%DocumentFonts: 
%%BoundingBox: 0 0 432 194
%%Orientation: Landscape
%%EndComments
/gnudict 256 dict def
gnudict begin
/Color false def
/Solid false def
/gnulinewidth 5.000 def
/userlinewidth gnulinewidth def
/vshift -33 def
/dl {10 mul} def
/hpt_ 31.5 def
/vpt_ 31.5 def
/hpt hpt_ def
/vpt vpt_ def
/M {moveto} bind def
/L {lineto} bind def
/R {rmoveto} bind def
/V {rlineto} bind def
/vpt2 vpt 2 mul def
/hpt2 hpt 2 mul def
/Lshow { currentpoint stroke M
  0 vshift R show } def
/Rshow { currentpoint stroke M
  dup stringwidth pop neg vshift R show } def
/Cshow { currentpoint stroke M
  dup stringwidth pop -2 div vshift R show } def
/UP { dup vpt_ mul /vpt exch def hpt_ mul /hpt exch def
  /hpt2 hpt 2 mul def /vpt2 vpt 2 mul def } def
/DL { Color {setrgbcolor Solid {pop []} if 0 setdash }
 {pop pop pop Solid {pop []} if 0 setdash} ifelse } def
/BL { stroke userlinewidth 2 mul setlinewidth } def
/AL { stroke userlinewidth 2 div setlinewidth } def
/UL { dup gnulinewidth mul /userlinewidth exch def
      10 mul /udl exch def } def
/PL { stroke userlinewidth setlinewidth } def
/LTb { BL [] 0 0 0 DL } def
/LTa { AL [1 udl mul 2 udl mul] 0 setdash 0 0 0 setrgbcolor } def
/LT0 { PL [] 1 0 0 DL } def
/LT1 { PL [4 dl 2 dl] 0 1 0 DL } def
/LT2 { PL [2 dl 3 dl] 0 0 1 DL } def
/LT3 { PL [1 dl 1.5 dl] 1 0 1 DL } def
/LT4 { PL [5 dl 2 dl 1 dl 2 dl] 0 1 1 DL } def
/LT5 { PL [4 dl 3 dl 1 dl 3 dl] 1 1 0 DL } def
/LT6 { PL [2 dl 2 dl 2 dl 4 dl] 0 0 0 DL } def
/LT7 { PL [2 dl 2 dl 2 dl 2 dl 2 dl 4 dl] 1 0.3 0 DL } def
/LT8 { PL [2 dl 2 dl 2 dl 2 dl 2 dl 2 dl 2 dl 4 dl] 0.5 0.5 0.5 DL } def
/Pnt { stroke [] 0 setdash
   gsave 1 setlinecap M 0 0 V stroke grestore } def
/Dia { stroke [] 0 setdash 2 copy vpt add M
  hpt neg vpt neg V hpt vpt neg V
  hpt vpt V hpt neg vpt V closepath stroke
  Pnt } def
/Pls { stroke [] 0 setdash vpt sub M 0 vpt2 V
  currentpoint stroke M
  hpt neg vpt neg R hpt2 0 V stroke
  } def
/Box { stroke [] 0 setdash 2 copy exch hpt sub exch vpt add M
  0 vpt2 neg V hpt2 0 V 0 vpt2 V
  hpt2 neg 0 V closepath stroke
  Pnt } def
/Crs { stroke [] 0 setdash exch hpt sub exch vpt add M
  hpt2 vpt2 neg V currentpoint stroke M
  hpt2 neg 0 R hpt2 vpt2 V stroke } def
/TriU { stroke [] 0 setdash 2 copy vpt 1.12 mul add M
  hpt neg vpt -1.62 mul V
  hpt 2 mul 0 V
  hpt neg vpt 1.62 mul V closepath stroke
  Pnt  } def
/Star { 2 copy Pls Crs } def
/BoxF { stroke [] 0 setdash exch hpt sub exch vpt add M
  0 vpt2 neg V  hpt2 0 V  0 vpt2 V
  hpt2 neg 0 V  closepath fill } def
/TriUF { stroke [] 0 setdash vpt 1.12 mul add M
  hpt neg vpt -1.62 mul V
  hpt 2 mul 0 V
  hpt neg vpt 1.62 mul V closepath fill } def
/TriD { stroke [] 0 setdash 2 copy vpt 1.12 mul sub M
  hpt neg vpt 1.62 mul V
  hpt 2 mul 0 V
  hpt neg vpt -1.62 mul V closepath stroke
  Pnt  } def
/TriDF { stroke [] 0 setdash vpt 1.12 mul sub M
  hpt neg vpt 1.62 mul V
  hpt 2 mul 0 V
  hpt neg vpt -1.62 mul V closepath fill} def
/DiaF { stroke [] 0 setdash vpt add M
  hpt neg vpt neg V hpt vpt neg V
  hpt vpt V hpt neg vpt V closepath fill } def
/Pent { stroke [] 0 setdash 2 copy gsave
  translate 0 hpt M 4 {72 rotate 0 hpt L} repeat
  closepath stroke grestore Pnt } def
/PentF { stroke [] 0 setdash gsave
  translate 0 hpt M 4 {72 rotate 0 hpt L} repeat
  closepath fill grestore } def
/Circle { stroke [] 0 setdash 2 copy
  hpt 0 360 arc stroke Pnt } def
/CircleF { stroke [] 0 setdash hpt 0 360 arc fill } def
/C0 { BL [] 0 setdash 2 copy moveto vpt 90 450  arc } bind def
/C1 { BL [] 0 setdash 2 copy        moveto
       2 copy  vpt 0 90 arc closepath fill
               vpt 0 360 arc closepath } bind def
/C2 { BL [] 0 setdash 2 copy moveto
       2 copy  vpt 90 180 arc closepath fill
               vpt 0 360 arc closepath } bind def
/C3 { BL [] 0 setdash 2 copy moveto
       2 copy  vpt 0 180 arc closepath fill
               vpt 0 360 arc closepath } bind def
/C4 { BL [] 0 setdash 2 copy moveto
       2 copy  vpt 180 270 arc closepath fill
               vpt 0 360 arc closepath } bind def
/C5 { BL [] 0 setdash 2 copy moveto
       2 copy  vpt 0 90 arc
       2 copy moveto
       2 copy  vpt 180 270 arc closepath fill
               vpt 0 360 arc } bind def
/C6 { BL [] 0 setdash 2 copy moveto
      2 copy  vpt 90 270 arc closepath fill
              vpt 0 360 arc closepath } bind def
/C7 { BL [] 0 setdash 2 copy moveto
      2 copy  vpt 0 270 arc closepath fill
              vpt 0 360 arc closepath } bind def
/C8 { BL [] 0 setdash 2 copy moveto
      2 copy vpt 270 360 arc closepath fill
              vpt 0 360 arc closepath } bind def
/C9 { BL [] 0 setdash 2 copy moveto
      2 copy  vpt 270 450 arc closepath fill
              vpt 0 360 arc closepath } bind def
/C10 { BL [] 0 setdash 2 copy 2 copy moveto vpt 270 360 arc closepath fill
       2 copy moveto
       2 copy vpt 90 180 arc closepath fill
               vpt 0 360 arc closepath } bind def
/C11 { BL [] 0 setdash 2 copy moveto
       2 copy  vpt 0 180 arc closepath fill
       2 copy moveto
       2 copy  vpt 270 360 arc closepath fill
               vpt 0 360 arc closepath } bind def
/C12 { BL [] 0 setdash 2 copy moveto
       2 copy  vpt 180 360 arc closepath fill
               vpt 0 360 arc closepath } bind def
/C13 { BL [] 0 setdash  2 copy moveto
       2 copy  vpt 0 90 arc closepath fill
       2 copy moveto
       2 copy  vpt 180 360 arc closepath fill
               vpt 0 360 arc closepath } bind def
/C14 { BL [] 0 setdash 2 copy moveto
       2 copy  vpt 90 360 arc closepath fill
               vpt 0 360 arc } bind def
/C15 { BL [] 0 setdash 2 copy vpt 0 360 arc closepath fill
               vpt 0 360 arc closepath } bind def
/Rec   { newpath 4 2 roll moveto 1 index 0 rlineto 0 exch rlineto
       neg 0 rlineto closepath } bind def
/Square { dup Rec } bind def
/Bsquare { vpt sub exch vpt sub exch vpt2 Square } bind def
/S0 { BL [] 0 setdash 2 copy moveto 0 vpt rlineto BL Bsquare } bind def
/S1 { BL [] 0 setdash 2 copy vpt Square fill Bsquare } bind def
/S2 { BL [] 0 setdash 2 copy exch vpt sub exch vpt Square fill Bsquare } bind def
/S3 { BL [] 0 setdash 2 copy exch vpt sub exch vpt2 vpt Rec fill Bsquare } bind def
/S4 { BL [] 0 setdash 2 copy exch vpt sub exch vpt sub vpt Square fill Bsquare } bind def
/S5 { BL [] 0 setdash 2 copy 2 copy vpt Square fill
       exch vpt sub exch vpt sub vpt Square fill Bsquare } bind def
/S6 { BL [] 0 setdash 2 copy exch vpt sub exch vpt sub vpt vpt2 Rec fill Bsquare } bind def
/S7 { BL [] 0 setdash 2 copy exch vpt sub exch vpt sub vpt vpt2 Rec fill
       2 copy vpt Square fill
       Bsquare } bind def
/S8 { BL [] 0 setdash 2 copy vpt sub vpt Square fill Bsquare } bind def
/S9 { BL [] 0 setdash 2 copy vpt sub vpt vpt2 Rec fill Bsquare } bind def
/S10 { BL [] 0 setdash 2 copy vpt sub vpt Square fill 2 copy exch vpt sub exch vpt Square fill
       Bsquare } bind def
/S11 { BL [] 0 setdash 2 copy vpt sub vpt Square fill 2 copy exch vpt sub exch vpt2 vpt Rec fill
       Bsquare } bind def
/S12 { BL [] 0 setdash 2 copy exch vpt sub exch vpt sub vpt2 vpt Rec fill Bsquare } bind def
/S13 { BL [] 0 setdash 2 copy exch vpt sub exch vpt sub vpt2 vpt Rec fill
       2 copy vpt Square fill Bsquare } bind def
/S14 { BL [] 0 setdash 2 copy exch vpt sub exch vpt sub vpt2 vpt Rec fill
       2 copy exch vpt sub exch vpt Square fill Bsquare } bind def
/S15 { BL [] 0 setdash 2 copy Bsquare fill Bsquare } bind def
/D0 { gsave translate 45 rotate 0 0 S0 stroke grestore } bind def
/D1 { gsave translate 45 rotate 0 0 S1 stroke grestore } bind def
/D2 { gsave translate 45 rotate 0 0 S2 stroke grestore } bind def
/D3 { gsave translate 45 rotate 0 0 S3 stroke grestore } bind def
/D4 { gsave translate 45 rotate 0 0 S4 stroke grestore } bind def
/D5 { gsave translate 45 rotate 0 0 S5 stroke grestore } bind def
/D6 { gsave translate 45 rotate 0 0 S6 stroke grestore } bind def
/D7 { gsave translate 45 rotate 0 0 S7 stroke grestore } bind def
/D8 { gsave translate 45 rotate 0 0 S8 stroke grestore } bind def
/D9 { gsave translate 45 rotate 0 0 S9 stroke grestore } bind def
/D10 { gsave translate 45 rotate 0 0 S10 stroke grestore } bind def
/D11 { gsave translate 45 rotate 0 0 S11 stroke grestore } bind def
/D12 { gsave translate 45 rotate 0 0 S12 stroke grestore } bind def
/D13 { gsave translate 45 rotate 0 0 S13 stroke grestore } bind def
/D14 { gsave translate 45 rotate 0 0 S14 stroke grestore } bind def
/D15 { gsave translate 45 rotate 0 0 S15 stroke grestore } bind def
/DiaE { stroke [] 0 setdash vpt add M
  hpt neg vpt neg V hpt vpt neg V
  hpt vpt V hpt neg vpt V closepath stroke } def
/BoxE { stroke [] 0 setdash exch hpt sub exch vpt add M
  0 vpt2 neg V hpt2 0 V 0 vpt2 V
  hpt2 neg 0 V closepath stroke } def
/TriUE { stroke [] 0 setdash vpt 1.12 mul add M
  hpt neg vpt -1.62 mul V
  hpt 2 mul 0 V
  hpt neg vpt 1.62 mul V closepath stroke } def
/TriDE { stroke [] 0 setdash vpt 1.12 mul sub M
  hpt neg vpt 1.62 mul V
  hpt 2 mul 0 V
  hpt neg vpt -1.62 mul V closepath stroke } def
/PentE { stroke [] 0 setdash gsave
  translate 0 hpt M 4 {72 rotate 0 hpt L} repeat
  closepath stroke grestore } def
/CircE { stroke [] 0 setdash 
  hpt 0 360 arc stroke } def
/Opaque { gsave closepath 1 setgray fill grestore 0 setgray closepath } def
/DiaW { stroke [] 0 setdash vpt add M
  hpt neg vpt neg V hpt vpt neg V
  hpt vpt V hpt neg vpt V Opaque stroke } def
/BoxW { stroke [] 0 setdash exch hpt sub exch vpt add M
  0 vpt2 neg V hpt2 0 V 0 vpt2 V
  hpt2 neg 0 V Opaque stroke } def
/TriUW { stroke [] 0 setdash vpt 1.12 mul add M
  hpt neg vpt -1.62 mul V
  hpt 2 mul 0 V
  hpt neg vpt 1.62 mul V Opaque stroke } def
/TriDW { stroke [] 0 setdash vpt 1.12 mul sub M
  hpt neg vpt 1.62 mul V
  hpt 2 mul 0 V
  hpt neg vpt -1.62 mul V Opaque stroke } def
/PentW { stroke [] 0 setdash gsave
  translate 0 hpt M 4 {72 rotate 0 hpt L} repeat
  Opaque stroke grestore } def
/CircW { stroke [] 0 setdash 
  hpt 0 360 arc Opaque stroke } def
/BoxFill { gsave Rec 1 setgray fill grestore } def
/Symbol-Oblique /Symbol findfont [1 0 .167 1 0 0] makefont
dup length dict begin {1 index /FID eq {pop pop} {def} ifelse} forall
currentdict end definefont
end
}}%
\begin{picture}(4320,1943)(0,0)%
{\GNUPLOTspecial{"
gnudict begin
gsave
0 0 translate
0.100 0.100 scale
0 setgray
newpath
1.000 UL
LTb
500 300 M
63 0 V
3607 0 R
-63 0 V
500 382 M
31 0 V
3639 0 R
-31 0 V
500 491 M
31 0 V
3639 0 R
-31 0 V
500 547 M
31 0 V
3639 0 R
-31 0 V
500 574 M
63 0 V
3607 0 R
-63 0 V
500 656 M
31 0 V
3639 0 R
-31 0 V
500 765 M
31 0 V
3639 0 R
-31 0 V
500 821 M
31 0 V
3639 0 R
-31 0 V
500 848 M
63 0 V
3607 0 R
-63 0 V
500 930 M
31 0 V
3639 0 R
-31 0 V
500 1039 M
31 0 V
3639 0 R
-31 0 V
500 1095 M
31 0 V
3639 0 R
-31 0 V
500 1121 M
63 0 V
3607 0 R
-63 0 V
500 1204 M
31 0 V
3639 0 R
-31 0 V
500 1313 M
31 0 V
3639 0 R
-31 0 V
500 1369 M
31 0 V
3639 0 R
-31 0 V
500 1395 M
63 0 V
3607 0 R
-63 0 V
500 1478 M
31 0 V
3639 0 R
-31 0 V
500 1587 M
31 0 V
3639 0 R
-31 0 V
500 1643 M
31 0 V
3639 0 R
-31 0 V
500 1669 M
63 0 V
3607 0 R
-63 0 V
500 1752 M
31 0 V
3639 0 R
-31 0 V
500 1861 M
31 0 V
3639 0 R
-31 0 V
500 1916 M
31 0 V
3639 0 R
-31 0 V
500 1943 M
63 0 V
3607 0 R
-63 0 V
500 300 M
0 63 V
0 1580 R
0 -63 V
868 300 M
0 31 V
0 1612 R
0 -31 V
1084 300 M
0 31 V
0 1612 R
0 -31 V
1237 300 M
0 31 V
0 1612 R
0 -31 V
1355 300 M
0 31 V
0 1612 R
0 -31 V
1452 300 M
0 31 V
0 1612 R
0 -31 V
1534 300 M
0 31 V
0 1612 R
0 -31 V
1605 300 M
0 31 V
0 1612 R
0 -31 V
1667 300 M
0 31 V
0 1612 R
0 -31 V
1723 300 M
0 63 V
0 1580 R
0 -63 V
2092 300 M
0 31 V
0 1612 R
0 -31 V
2307 300 M
0 31 V
0 1612 R
0 -31 V
2460 300 M
0 31 V
0 1612 R
0 -31 V
2578 300 M
0 31 V
0 1612 R
0 -31 V
2675 300 M
0 31 V
0 1612 R
0 -31 V
2757 300 M
0 31 V
0 1612 R
0 -31 V
2828 300 M
0 31 V
0 1612 R
0 -31 V
2891 300 M
0 31 V
0 1612 R
0 -31 V
2947 300 M
0 63 V
0 1580 R
0 -63 V
3315 300 M
0 31 V
0 1612 R
0 -31 V
3530 300 M
0 31 V
0 1612 R
0 -31 V
3683 300 M
0 31 V
0 1612 R
0 -31 V
3802 300 M
0 31 V
0 1612 R
0 -31 V
3899 300 M
0 31 V
0 1612 R
0 -31 V
3981 300 M
0 31 V
0 1612 R
0 -31 V
4051 300 M
0 31 V
0 1612 R
0 -31 V
4114 300 M
0 31 V
0 1612 R
0 -31 V
4170 300 M
0 63 V
0 1580 R
0 -63 V
1.000 UL
LTb
500 300 M
3670 0 V
0 1643 V
-3670 0 V
500 300 L
1.000 UL
LT0
600 1830 M
263 0 V
4170 848 M
-1 0 V
-1 0 V
-1 0 V
-1 0 V
-1 0 V
-2 0 V
-2 0 V
-3 0 V
-5 0 V
-6 0 V
-9 0 V
-12 0 V
-16 0 V
-24 0 V
-34 0 V
-49 0 V
-77 0 V
-126 0 V
-161 0 V
-164 0 V
-164 0 V
-164 0 V
-164 0 V
-164 0 V
-163 0 V
-164 0 V
-164 0 V
-164 0 V
-164 0 V
-164 0 V
-164 0 V
-164 0 V
-164 0 V
-164 0 V
-164 0 V
-164 0 V
-164 0 V
-164 0 V
-24 0 V
1.000 UL
LT1
600 1730 M
263 0 V
3277 213 R
-2 -30 V
-12 -132 V
-16 -132 V
-24 -134 V
-34 -137 V
-49 -143 V
-77 -149 V
3800 944 L
3639 872 L
3475 857 L
-164 9 V
-164 15 V
-164 15 V
-164 13 V
-163 11 V
-164 9 V
-164 7 V
-164 5 V
-164 4 V
-164 3 V
-164 3 V
-164 2 V
-164 2 V
-164 2 V
-164 2 V
-164 2 V
-164 2 V
-164 0 V
-24 0 V
1.000 UL
LT2
600 1630 M
263 0 V
3297 313 R
-2 -86 V
-5 -99 V
-6 -96 V
-9 -92 V
-12 -87 V
-16 -79 V
-24 -69 V
-34 -57 V
-49 -50 V
-77 -54 V
-126 -70 V
-161 -68 V
3475 981 L
3311 938 L
3147 902 L
2983 871 L
2819 845 L
2656 821 L
2492 800 L
2328 779 L
2164 758 L
2000 739 L
1836 719 L
1672 701 L
1508 683 L
1344 666 L
1180 650 L
1016 634 L
852 618 L
688 603 L
524 587 L
-24 -2 V
1.000 UL
LT3
600 1530 M
263 0 V
3298 413 R
-3 -84 V
-5 -89 V
-6 -87 V
-9 -86 V
-12 -85 V
-16 -86 V
-24 -88 V
-34 -92 V
-49 -98 V
-77 -106 V
3800 931 L
3639 855 L
3475 820 L
-164 -8 V
-164 7 V
-164 14 V
-164 15 V
-163 15 V
-164 12 V
-164 10 V
-164 8 V
-164 6 V
-164 4 V
-164 3 V
-164 3 V
-164 3 V
-164 2 V
-164 2 V
-164 3 V
-164 2 V
-164 0 V
-24 1 V
1.000 UL
LT4
600 1430 M
263 0 V
3297 513 R
-2 -72 V
-5 -88 V
-6 -86 V
-9 -85 V
-12 -84 V
-16 -86 V
-24 -87 V
-34 -91 V
-49 -99 V
-77 -107 V
3800 944 L
3639 865 L
3475 826 L
3311 815 L
-164 6 V
-164 12 V
-164 15 V
-163 15 V
-164 12 V
-164 9 V
-164 8 V
-164 5 V
-164 5 V
-164 3 V
-164 3 V
-164 2 V
-164 3 V
-164 2 V
-164 2 V
-164 2 V
-164 1 V
-24 0 V
1.000 UL
LT5
600 1330 M
263 0 V
3297 613 R
-2 -75 V
-5 -88 V
-6 -86 V
-9 -85 V
-12 -85 V
-16 -85 V
-24 -88 V
-34 -90 V
-49 -97 V
-77 -104 V
3800 955 L
3639 889 L
3475 861 L
-164 -3 V
-164 11 V
-164 17 V
-164 20 V
-163 17 V
-164 15 V
-164 12 V
-164 9 V
-164 7 V
-164 5 V
-164 5 V
-164 3 V
-164 3 V
-164 3 V
-164 3 V
-164 2 V
-164 3 V
-164 1 V
-24 1 V
1.000 UL
LT6
600 1230 M
263 0 V
3297 713 R
-2 -72 V
-5 -87 V
-6 -86 V
-9 -84 V
-12 -85 V
-16 -85 V
-24 -87 V
-34 -92 V
-49 -101 V
-77 -115 V
3800 915 L
3639 811 L
3475 748 L
3311 710 L
3147 687 L
2983 670 L
2819 657 L
2656 646 L
2492 635 L
2328 624 L
2164 612 L
2000 599 L
1836 585 L
1672 571 L
1508 557 L
1344 543 L
1180 529 L
1016 516 L
852 502 L
688 489 L
524 474 L
-24 -1 V
stroke
grestore
end
showpage
}}%
\put(913,1230){\makebox(0,0)[l]{$u_L \rightarrow \nu_\tau$}}%
\put(913,1330){\makebox(0,0)[l]{$u_L \rightarrow \nu_\mu$}}%
\put(913,1430){\makebox(0,0)[l]{$u_L \rightarrow \nu_e$}}%
\put(913,1530){\makebox(0,0)[l]{$u_L \rightarrow e$}}%
\put(913,1630){\makebox(0,0)[l]{$u_L \rightarrow LSP$}}%
\put(913,1730){\makebox(0,0)[l]{$u_L \rightarrow \gamma$}}%
\put(913,1830){\makebox(0,0)[l]{$u_L \rightarrow p$}}%
\put(2335,50){\makebox(0,0){$x$}}%
\put(100,1121){%
\special{ps: gsave currentpoint currentpoint translate
270 rotate neg exch neg exch translate}%
\makebox(0,0)[b]{\shortstack{$D_{u_L}^i(x,M_X)/D_{u_L}^p(x,M_X)$}}%
\special{ps: currentpoint grestore moveto}%
}%
\put(4170,200){\makebox(0,0){1}}%
\put(2947,200){\makebox(0,0){0.1}}%
\put(1723,200){\makebox(0,0){0.01}}%
\put(500,200){\makebox(0,0){0.001}}%
\put(450,1943){\makebox(0,0)[r]{10000}}%
\put(450,1669){\makebox(0,0)[r]{1000}}%
\put(450,1395){\makebox(0,0)[r]{100}}%
\put(450,1121){\makebox(0,0)[r]{10}}%
\put(450,848){\makebox(0,0)[r]{1}}%
\put(450,574){\makebox(0,0)[r]{0.1}}%
\put(450,300){\makebox(0,0)[r]{0.01}}%
\end{picture}%
\endgroup
 

%% file: Low_G_ratio_eL_.tex
% GNUPLOT: LaTeX picture with Postscript
\begingroup%
  \makeatletter%
  \newcommand{\GNUPLOTspecial}{%
    \@sanitize\catcode`\%=14\relax\special}%
  \setlength{\unitlength}{0.1bp}%
{\GNUPLOTspecial{!
%!PS-Adobe-2.0 EPSF-2.0
%%Title: /a/srv-1.cip/h/t30i/dre/barbot/Latex/Talk_Susy02/Low_G_ratio_eL_.tex
%%Creator: gnuplot 3.7 patchlevel 1
%%CreationDate: Thu Oct 17 15:38:19 2002
%%DocumentFonts: 
%%BoundingBox: 0 0 432 194
%%Orientation: Landscape
%%EndComments
/gnudict 256 dict def
gnudict begin
/Color false def
/Solid false def
/gnulinewidth 5.000 def
/userlinewidth gnulinewidth def
/vshift -33 def
/dl {10 mul} def
/hpt_ 31.5 def
/vpt_ 31.5 def
/hpt hpt_ def
/vpt vpt_ def
/M {moveto} bind def
/L {lineto} bind def
/R {rmoveto} bind def
/V {rlineto} bind def
/vpt2 vpt 2 mul def
/hpt2 hpt 2 mul def
/Lshow { currentpoint stroke M
  0 vshift R show } def
/Rshow { currentpoint stroke M
  dup stringwidth pop neg vshift R show } def
/Cshow { currentpoint stroke M
  dup stringwidth pop -2 div vshift R show } def
/UP { dup vpt_ mul /vpt exch def hpt_ mul /hpt exch def
  /hpt2 hpt 2 mul def /vpt2 vpt 2 mul def } def
/DL { Color {setrgbcolor Solid {pop []} if 0 setdash }
 {pop pop pop Solid {pop []} if 0 setdash} ifelse } def
/BL { stroke userlinewidth 2 mul setlinewidth } def
/AL { stroke userlinewidth 2 div setlinewidth } def
/UL { dup gnulinewidth mul /userlinewidth exch def
      10 mul /udl exch def } def
/PL { stroke userlinewidth setlinewidth } def
/LTb { BL [] 0 0 0 DL } def
/LTa { AL [1 udl mul 2 udl mul] 0 setdash 0 0 0 setrgbcolor } def
/LT0 { PL [] 1 0 0 DL } def
/LT1 { PL [4 dl 2 dl] 0 1 0 DL } def
/LT2 { PL [2 dl 3 dl] 0 0 1 DL } def
/LT3 { PL [1 dl 1.5 dl] 1 0 1 DL } def
/LT4 { PL [5 dl 2 dl 1 dl 2 dl] 0 1 1 DL } def
/LT5 { PL [4 dl 3 dl 1 dl 3 dl] 1 1 0 DL } def
/LT6 { PL [2 dl 2 dl 2 dl 4 dl] 0 0 0 DL } def
/LT7 { PL [2 dl 2 dl 2 dl 2 dl 2 dl 4 dl] 1 0.3 0 DL } def
/LT8 { PL [2 dl 2 dl 2 dl 2 dl 2 dl 2 dl 2 dl 4 dl] 0.5 0.5 0.5 DL } def
/Pnt { stroke [] 0 setdash
   gsave 1 setlinecap M 0 0 V stroke grestore } def
/Dia { stroke [] 0 setdash 2 copy vpt add M
  hpt neg vpt neg V hpt vpt neg V
  hpt vpt V hpt neg vpt V closepath stroke
  Pnt } def
/Pls { stroke [] 0 setdash vpt sub M 0 vpt2 V
  currentpoint stroke M
  hpt neg vpt neg R hpt2 0 V stroke
  } def
/Box { stroke [] 0 setdash 2 copy exch hpt sub exch vpt add M
  0 vpt2 neg V hpt2 0 V 0 vpt2 V
  hpt2 neg 0 V closepath stroke
  Pnt } def
/Crs { stroke [] 0 setdash exch hpt sub exch vpt add M
  hpt2 vpt2 neg V currentpoint stroke M
  hpt2 neg 0 R hpt2 vpt2 V stroke } def
/TriU { stroke [] 0 setdash 2 copy vpt 1.12 mul add M
  hpt neg vpt -1.62 mul V
  hpt 2 mul 0 V
  hpt neg vpt 1.62 mul V closepath stroke
  Pnt  } def
/Star { 2 copy Pls Crs } def
/BoxF { stroke [] 0 setdash exch hpt sub exch vpt add M
  0 vpt2 neg V  hpt2 0 V  0 vpt2 V
  hpt2 neg 0 V  closepath fill } def
/TriUF { stroke [] 0 setdash vpt 1.12 mul add M
  hpt neg vpt -1.62 mul V
  hpt 2 mul 0 V
  hpt neg vpt 1.62 mul V closepath fill } def
/TriD { stroke [] 0 setdash 2 copy vpt 1.12 mul sub M
  hpt neg vpt 1.62 mul V
  hpt 2 mul 0 V
  hpt neg vpt -1.62 mul V closepath stroke
  Pnt  } def
/TriDF { stroke [] 0 setdash vpt 1.12 mul sub M
  hpt neg vpt 1.62 mul V
  hpt 2 mul 0 V
  hpt neg vpt -1.62 mul V closepath fill} def
/DiaF { stroke [] 0 setdash vpt add M
  hpt neg vpt neg V hpt vpt neg V
  hpt vpt V hpt neg vpt V closepath fill } def
/Pent { stroke [] 0 setdash 2 copy gsave
  translate 0 hpt M 4 {72 rotate 0 hpt L} repeat
  closepath stroke grestore Pnt } def
/PentF { stroke [] 0 setdash gsave
  translate 0 hpt M 4 {72 rotate 0 hpt L} repeat
  closepath fill grestore } def
/Circle { stroke [] 0 setdash 2 copy
  hpt 0 360 arc stroke Pnt } def
/CircleF { stroke [] 0 setdash hpt 0 360 arc fill } def
/C0 { BL [] 0 setdash 2 copy moveto vpt 90 450  arc } bind def
/C1 { BL [] 0 setdash 2 copy        moveto
       2 copy  vpt 0 90 arc closepath fill
               vpt 0 360 arc closepath } bind def
/C2 { BL [] 0 setdash 2 copy moveto
       2 copy  vpt 90 180 arc closepath fill
               vpt 0 360 arc closepath } bind def
/C3 { BL [] 0 setdash 2 copy moveto
       2 copy  vpt 0 180 arc closepath fill
               vpt 0 360 arc closepath } bind def
/C4 { BL [] 0 setdash 2 copy moveto
       2 copy  vpt 180 270 arc closepath fill
               vpt 0 360 arc closepath } bind def
/C5 { BL [] 0 setdash 2 copy moveto
       2 copy  vpt 0 90 arc
       2 copy moveto
       2 copy  vpt 180 270 arc closepath fill
               vpt 0 360 arc } bind def
/C6 { BL [] 0 setdash 2 copy moveto
      2 copy  vpt 90 270 arc closepath fill
              vpt 0 360 arc closepath } bind def
/C7 { BL [] 0 setdash 2 copy moveto
      2 copy  vpt 0 270 arc closepath fill
              vpt 0 360 arc closepath } bind def
/C8 { BL [] 0 setdash 2 copy moveto
      2 copy vpt 270 360 arc closepath fill
              vpt 0 360 arc closepath } bind def
/C9 { BL [] 0 setdash 2 copy moveto
      2 copy  vpt 270 450 arc closepath fill
              vpt 0 360 arc closepath } bind def
/C10 { BL [] 0 setdash 2 copy 2 copy moveto vpt 270 360 arc closepath fill
       2 copy moveto
       2 copy vpt 90 180 arc closepath fill
               vpt 0 360 arc closepath } bind def
/C11 { BL [] 0 setdash 2 copy moveto
       2 copy  vpt 0 180 arc closepath fill
       2 copy moveto
       2 copy  vpt 270 360 arc closepath fill
               vpt 0 360 arc closepath } bind def
/C12 { BL [] 0 setdash 2 copy moveto
       2 copy  vpt 180 360 arc closepath fill
               vpt 0 360 arc closepath } bind def
/C13 { BL [] 0 setdash  2 copy moveto
       2 copy  vpt 0 90 arc closepath fill
       2 copy moveto
       2 copy  vpt 180 360 arc closepath fill
               vpt 0 360 arc closepath } bind def
/C14 { BL [] 0 setdash 2 copy moveto
       2 copy  vpt 90 360 arc closepath fill
               vpt 0 360 arc } bind def
/C15 { BL [] 0 setdash 2 copy vpt 0 360 arc closepath fill
               vpt 0 360 arc closepath } bind def
/Rec   { newpath 4 2 roll moveto 1 index 0 rlineto 0 exch rlineto
       neg 0 rlineto closepath } bind def
/Square { dup Rec } bind def
/Bsquare { vpt sub exch vpt sub exch vpt2 Square } bind def
/S0 { BL [] 0 setdash 2 copy moveto 0 vpt rlineto BL Bsquare } bind def
/S1 { BL [] 0 setdash 2 copy vpt Square fill Bsquare } bind def
/S2 { BL [] 0 setdash 2 copy exch vpt sub exch vpt Square fill Bsquare } bind def
/S3 { BL [] 0 setdash 2 copy exch vpt sub exch vpt2 vpt Rec fill Bsquare } bind def
/S4 { BL [] 0 setdash 2 copy exch vpt sub exch vpt sub vpt Square fill Bsquare } bind def
/S5 { BL [] 0 setdash 2 copy 2 copy vpt Square fill
       exch vpt sub exch vpt sub vpt Square fill Bsquare } bind def
/S6 { BL [] 0 setdash 2 copy exch vpt sub exch vpt sub vpt vpt2 Rec fill Bsquare } bind def
/S7 { BL [] 0 setdash 2 copy exch vpt sub exch vpt sub vpt vpt2 Rec fill
       2 copy vpt Square fill
       Bsquare } bind def
/S8 { BL [] 0 setdash 2 copy vpt sub vpt Square fill Bsquare } bind def
/S9 { BL [] 0 setdash 2 copy vpt sub vpt vpt2 Rec fill Bsquare } bind def
/S10 { BL [] 0 setdash 2 copy vpt sub vpt Square fill 2 copy exch vpt sub exch vpt Square fill
       Bsquare } bind def
/S11 { BL [] 0 setdash 2 copy vpt sub vpt Square fill 2 copy exch vpt sub exch vpt2 vpt Rec fill
       Bsquare } bind def
/S12 { BL [] 0 setdash 2 copy exch vpt sub exch vpt sub vpt2 vpt Rec fill Bsquare } bind def
/S13 { BL [] 0 setdash 2 copy exch vpt sub exch vpt sub vpt2 vpt Rec fill
       2 copy vpt Square fill Bsquare } bind def
/S14 { BL [] 0 setdash 2 copy exch vpt sub exch vpt sub vpt2 vpt Rec fill
       2 copy exch vpt sub exch vpt Square fill Bsquare } bind def
/S15 { BL [] 0 setdash 2 copy Bsquare fill Bsquare } bind def
/D0 { gsave translate 45 rotate 0 0 S0 stroke grestore } bind def
/D1 { gsave translate 45 rotate 0 0 S1 stroke grestore } bind def
/D2 { gsave translate 45 rotate 0 0 S2 stroke grestore } bind def
/D3 { gsave translate 45 rotate 0 0 S3 stroke grestore } bind def
/D4 { gsave translate 45 rotate 0 0 S4 stroke grestore } bind def
/D5 { gsave translate 45 rotate 0 0 S5 stroke grestore } bind def
/D6 { gsave translate 45 rotate 0 0 S6 stroke grestore } bind def
/D7 { gsave translate 45 rotate 0 0 S7 stroke grestore } bind def
/D8 { gsave translate 45 rotate 0 0 S8 stroke grestore } bind def
/D9 { gsave translate 45 rotate 0 0 S9 stroke grestore } bind def
/D10 { gsave translate 45 rotate 0 0 S10 stroke grestore } bind def
/D11 { gsave translate 45 rotate 0 0 S11 stroke grestore } bind def
/D12 { gsave translate 45 rotate 0 0 S12 stroke grestore } bind def
/D13 { gsave translate 45 rotate 0 0 S13 stroke grestore } bind def
/D14 { gsave translate 45 rotate 0 0 S14 stroke grestore } bind def
/D15 { gsave translate 45 rotate 0 0 S15 stroke grestore } bind def
/DiaE { stroke [] 0 setdash vpt add M
  hpt neg vpt neg V hpt vpt neg V
  hpt vpt V hpt neg vpt V closepath stroke } def
/BoxE { stroke [] 0 setdash exch hpt sub exch vpt add M
  0 vpt2 neg V hpt2 0 V 0 vpt2 V
  hpt2 neg 0 V closepath stroke } def
/TriUE { stroke [] 0 setdash vpt 1.12 mul add M
  hpt neg vpt -1.62 mul V
  hpt 2 mul 0 V
  hpt neg vpt 1.62 mul V closepath stroke } def
/TriDE { stroke [] 0 setdash vpt 1.12 mul sub M
  hpt neg vpt 1.62 mul V
  hpt 2 mul 0 V
  hpt neg vpt -1.62 mul V closepath stroke } def
/PentE { stroke [] 0 setdash gsave
  translate 0 hpt M 4 {72 rotate 0 hpt L} repeat
  closepath stroke grestore } def
/CircE { stroke [] 0 setdash 
  hpt 0 360 arc stroke } def
/Opaque { gsave closepath 1 setgray fill grestore 0 setgray closepath } def
/DiaW { stroke [] 0 setdash vpt add M
  hpt neg vpt neg V hpt vpt neg V
  hpt vpt V hpt neg vpt V Opaque stroke } def
/BoxW { stroke [] 0 setdash exch hpt sub exch vpt add M
  0 vpt2 neg V hpt2 0 V 0 vpt2 V
  hpt2 neg 0 V Opaque stroke } def
/TriUW { stroke [] 0 setdash vpt 1.12 mul add M
  hpt neg vpt -1.62 mul V
  hpt 2 mul 0 V
  hpt neg vpt 1.62 mul V Opaque stroke } def
/TriDW { stroke [] 0 setdash vpt 1.12 mul sub M
  hpt neg vpt 1.62 mul V
  hpt 2 mul 0 V
  hpt neg vpt -1.62 mul V Opaque stroke } def
/PentW { stroke [] 0 setdash gsave
  translate 0 hpt M 4 {72 rotate 0 hpt L} repeat
  Opaque stroke grestore } def
/CircW { stroke [] 0 setdash 
  hpt 0 360 arc Opaque stroke } def
/BoxFill { gsave Rec 1 setgray fill grestore } def
/Symbol-Oblique /Symbol findfont [1 0 .167 1 0 0] makefont
dup length dict begin {1 index /FID eq {pop pop} {def} ifelse} forall
currentdict end definefont
end
}}%
\begin{picture}(4320,1943)(0,0)%
{\GNUPLOTspecial{"
gnudict begin
gsave
0 0 translate
0.100 0.100 scale
0 setgray
newpath
1.000 UL
LTb
500 300 M
63 0 V
3607 0 R
-63 0 V
500 382 M
31 0 V
3639 0 R
-31 0 V
500 491 M
31 0 V
3639 0 R
-31 0 V
500 547 M
31 0 V
3639 0 R
-31 0 V
500 574 M
63 0 V
3607 0 R
-63 0 V
500 656 M
31 0 V
3639 0 R
-31 0 V
500 765 M
31 0 V
3639 0 R
-31 0 V
500 821 M
31 0 V
3639 0 R
-31 0 V
500 848 M
63 0 V
3607 0 R
-63 0 V
500 930 M
31 0 V
3639 0 R
-31 0 V
500 1039 M
31 0 V
3639 0 R
-31 0 V
500 1095 M
31 0 V
3639 0 R
-31 0 V
500 1121 M
63 0 V
3607 0 R
-63 0 V
500 1204 M
31 0 V
3639 0 R
-31 0 V
500 1313 M
31 0 V
3639 0 R
-31 0 V
500 1369 M
31 0 V
3639 0 R
-31 0 V
500 1395 M
63 0 V
3607 0 R
-63 0 V
500 1478 M
31 0 V
3639 0 R
-31 0 V
500 1587 M
31 0 V
3639 0 R
-31 0 V
500 1643 M
31 0 V
3639 0 R
-31 0 V
500 1669 M
63 0 V
3607 0 R
-63 0 V
500 1752 M
31 0 V
3639 0 R
-31 0 V
500 1861 M
31 0 V
3639 0 R
-31 0 V
500 1916 M
31 0 V
3639 0 R
-31 0 V
500 1943 M
63 0 V
3607 0 R
-63 0 V
500 300 M
0 63 V
0 1580 R
0 -63 V
868 300 M
0 31 V
0 1612 R
0 -31 V
1084 300 M
0 31 V
0 1612 R
0 -31 V
1237 300 M
0 31 V
0 1612 R
0 -31 V
1355 300 M
0 31 V
0 1612 R
0 -31 V
1452 300 M
0 31 V
0 1612 R
0 -31 V
1534 300 M
0 31 V
0 1612 R
0 -31 V
1605 300 M
0 31 V
0 1612 R
0 -31 V
1667 300 M
0 31 V
0 1612 R
0 -31 V
1723 300 M
0 63 V
0 1580 R
0 -63 V
2092 300 M
0 31 V
0 1612 R
0 -31 V
2307 300 M
0 31 V
0 1612 R
0 -31 V
2460 300 M
0 31 V
0 1612 R
0 -31 V
2578 300 M
0 31 V
0 1612 R
0 -31 V
2675 300 M
0 31 V
0 1612 R
0 -31 V
2757 300 M
0 31 V
0 1612 R
0 -31 V
2828 300 M
0 31 V
0 1612 R
0 -31 V
2891 300 M
0 31 V
0 1612 R
0 -31 V
2947 300 M
0 63 V
0 1580 R
0 -63 V
3315 300 M
0 31 V
0 1612 R
0 -31 V
3530 300 M
0 31 V
0 1612 R
0 -31 V
3683 300 M
0 31 V
0 1612 R
0 -31 V
3802 300 M
0 31 V
0 1612 R
0 -31 V
3899 300 M
0 31 V
0 1612 R
0 -31 V
3981 300 M
0 31 V
0 1612 R
0 -31 V
4051 300 M
0 31 V
0 1612 R
0 -31 V
4114 300 M
0 31 V
0 1612 R
0 -31 V
4170 300 M
0 63 V
0 1580 R
0 -63 V
1.000 UL
LTb
500 300 M
3670 0 V
0 1643 V
-3670 0 V
500 300 L
1.000 UL
LT0
600 1830 M
263 0 V
4170 848 M
-1 0 V
-1 0 V
-1 0 V
-1 0 V
-1 0 V
-2 0 V
-2 0 V
-3 0 V
-5 0 V
-6 0 V
-9 0 V
-3 0 V
-3 0 V
-2 0 V
-1 0 V
-1 0 V
-1 0 V
-1 0 V
-1 0 V
-1 0 V
-1 0 V
-1 0 V
-1 0 V
-2 0 V
-2 0 V
-3 0 V
-3 0 V
-6 0 V
-8 0 V
-11 0 V
-34 0 V
-49 0 V
-77 0 V
-126 0 V
-161 0 V
-164 0 V
-164 0 V
-164 0 V
-164 0 V
-164 0 V
-163 0 V
-164 0 V
-164 0 V
-164 0 V
-164 0 V
-164 0 V
-164 0 V
-164 0 V
-164 0 V
-164 0 V
-164 0 V
-164 0 V
-164 0 V
-164 0 V
-24 0 V
1.000 UL
LT1
600 1730 M
263 0 V
2982 213 R
-45 -102 V
3639 1566 L
3475 1353 L
3311 1192 L
3147 1076 L
-164 -76 V
2819 955 L
2656 932 L
-164 -9 V
-164 -2 V
-164 3 V
-164 5 V
-164 6 V
-164 6 V
-164 6 V
-164 5 V
-164 5 V
-164 4 V
-164 4 V
-164 2 V
-164 3 V
-24 0 V
1.000 UL
LT2
600 1630 M
263 0 V
2755 313 R
3475 1734 L
3311 1528 L
3147 1358 L
2983 1217 L
2819 1097 L
2656 990 L
2492 890 L
2328 804 L
2164 740 L
2000 694 L
1836 661 L
1672 635 L
1508 612 L
1344 592 L
1180 574 L
1016 557 L
852 541 L
688 525 L
524 510 L
-24 -3 V
1.000 UL
LT3
600 1530 M
263 0 V
2826 413 R
-50 -94 V
3475 1615 L
3311 1434 L
3147 1291 L
2983 1178 L
-164 -90 V
-163 -69 V
2492 968 L
2328 933 L
2164 913 L
-164 -9 V
-164 -2 V
-164 2 V
-164 4 V
-164 6 V
-164 5 V
-164 5 V
-164 4 V
-164 4 V
-164 3 V
-24 0 V
1.000 UL
LT4
600 1430 M
263 0 V
2826 513 R
-50 -95 V
3475 1614 L
3311 1433 L
3147 1291 L
2983 1178 L
-164 -89 V
-163 -70 V
2492 968 L
2328 934 L
2164 913 L
-164 -9 V
-164 -3 V
-164 2 V
-164 5 V
-164 5 V
-164 5 V
-164 5 V
-164 4 V
-164 4 V
-164 3 V
-24 0 V
1.000 UL
LT5
600 1330 M
263 0 V
2826 613 R
-50 -95 V
3475 1614 L
3311 1433 L
3147 1292 L
2983 1180 L
-164 -86 V
-163 -63 V
2492 989 L
2328 965 L
2164 954 L
-164 -1 V
-164 5 V
-164 7 V
-164 8 V
-164 7 V
-164 7 V
-164 7 V
-164 5 V
-164 4 V
-164 4 V
-24 0 V
1.000 UL
LT6
600 1230 M
263 0 V
3077 713 R
-14 -30 V
3800 1753 L
3639 1567 L
3475 1401 L
3311 1263 L
3147 1151 L
-164 -92 V
2819 983 L
2656 919 L
2492 864 L
2328 816 L
2164 775 L
2000 739 L
1836 707 L
1672 680 L
1508 655 L
1344 632 L
1180 612 L
1016 593 L
852 575 L
688 558 L
524 541 L
-24 -2 V
stroke
grestore
end
showpage
}}%
\put(913,1230){\makebox(0,0)[l]{$\tilde{e}_L \rightarrow \nu_\tau$}}%
\put(913,1330){\makebox(0,0)[l]{$\tilde{e}_L \rightarrow \nu_\mu$}}%
\put(913,1430){\makebox(0,0)[l]{$\tilde{e}_L \rightarrow \nu_e$}}%
\put(913,1530){\makebox(0,0)[l]{$\tilde{e}_L \rightarrow e$}}%
\put(913,1630){\makebox(0,0)[l]{$\tilde{e}_L \rightarrow LSP$}}%
\put(913,1730){\makebox(0,0)[l]{$\tilde{e}_L \rightarrow \gamma$}}%
\put(913,1830){\makebox(0,0)[l]{$\tilde{e}_L \rightarrow p$}}%
\put(2335,50){\makebox(0,0){$x$}}%
\put(100,1121){%
\special{ps: gsave currentpoint currentpoint translate
270 rotate neg exch neg exch translate}%
\makebox(0,0)[b]{\shortstack{$D_{\tilde{e}_L}^i(x,M_X)/D_{\tilde{e}_L}^p(x,M_X)$}}%
\special{ps: currentpoint grestore moveto}%
}%
\put(4170,200){\makebox(0,0){1}}%
\put(2947,200){\makebox(0,0){0.1}}%
\put(1723,200){\makebox(0,0){0.01}}%
\put(500,200){\makebox(0,0){0.001}}%
\put(450,1943){\makebox(0,0)[r]{10000}}%
\put(450,1669){\makebox(0,0)[r]{1000}}%
\put(450,1395){\makebox(0,0)[r]{100}}%
\put(450,1121){\makebox(0,0)[r]{10}}%
\put(450,848){\makebox(0,0)[r]{1}}%
\put(450,574){\makebox(0,0)[r]{0.1}}%
\put(450,300){\makebox(0,0)[r]{0.01}}%
\end{picture}%
\endgroup
 